%% file: main.tex
\documentclass[sigconf]{acmart}
\pdfoutput=1
\usepackage{tcolorbox}
\tcbuselibrary{theorems}

\usepackage{enumerate}
\usepackage{algorithmic}
\usepackage{subcaption}
\usepackage{amsmath,amsfonts,amsthm}
\usepackage{algorithmic}
\usepackage{textcomp}
\usepackage{soul,xcolor, colortbl}
\usepackage{tikz}
\usepackage{enumitem}
\usepackage{pifont}
\usetikzlibrary{arrows.meta,
                chains,
                positioning,
                shapes.geometric,
                shapes.symbols
                }
\usepackage{comment}
\usepackage[english]{babel}
\usepackage{caption}
\usepackage{multirow}
\usepackage{multicol, blindtext}
\usepackage{diagbox}
\usepackage{hyperref}
\usepackage{multirow}
\usepackage{amsmath}
\usepackage{tabularx}
\usepackage{cancel}

\newcommand{\RN}[1]{%
  \textup{\uppercase\expandafter{\romannumeral#1}}%
}

\newcommand{\red}[1]{\textcolor{red}{#1}}

\newcommand{\green}[1]{\textcolor{green}{#1}}

\theoremstyle{definition}

\newtcbtheorem{mytheo}{Definition}%
{colbacktitle=black!75,colframe=black}{th}

\newlength\Origarrayrulewidth

\newcommand{\cmark}{\green{\ding{51}}}%
\newcommand{\xmark}{\red{\ding{55}}}%

\AtBeginDocument{%
  \providecommand\BibTeX{{%
    \normalfont B\kern-0.5em{\scshape i\kern-0.25em b}\kern-0.8em\TeX}}}

\setcopyright{acmcopyright}
\copyrightyear{2022}
\acmYear{2022}
\setcopyright{acmcopyright}\acmConference[ICSE '22]{44th International Conference on Software Engineering}{May 21--29, 2022}{Pittsburgh, PA, USA}
\acmBooktitle{44th International Conference on Software Engineering (ICSE '22), May 21--29, 2022, Pittsburgh, PA, USA}
\acmPrice{15.00}
\acmDOI{10.1145/3510003.3510109}
\acmISBN{978-1-4503-9221-1/22/05}

\begin{document}
\newpage

\title{If a Human Can See It, So Should Your System: Reliability Requirements for Machine Vision Components}

\author{Boyue Caroline Hu}
\email{boyue@cs.toronto.edu}
\affiliation{%
  \institution{University of Toronto}
  \city{Toronto}
  \state{Ontario}
  \country{Canada}
}

\author{Lina Marsso}
\email{lina.marsso@utoronto.ca}
\affiliation{%
  \institution{University of Toronto}
  \city{Toronto}
  \state{Ontario}
  \country{Canada}
}

\author{Krzysztof Czarnecki}
\email{kczarnec@gsd.uwaterloo.ca}
\affiliation{%
  \institution{University of Waterloo}
  \city{Waterloo}
  \state{Ontario}
  \country{Canada}
}

\author{Rick Salay}
\email{rsalay@gsd.uwaterloo.ca}
\affiliation{%
  \institution{University of Waterloo}
  \city{Waterloo}
  \state{Ontario}
  \country{Canada}
}

\author{Huakun Shen}
\email{huakun.shen@mail.utoronto.ca}
\affiliation{%
  \institution{University of Toronto}
  \city{Toronto}
  \state{Ontario}
  \country{Canada}
}
\author{Marsha Chechik}
\email{chechik@cs.toronto.edu}
\affiliation{%
  \institution{University of Toronto}
  \city{Toronto}
  \state{Ontario}
  \country{Canada}
}

\begin{abstract}
Machine Vision Components (MVC) are becoming safety-critical. Assuring their quality, including safety, is essential for their successful deployment. Assurance relies on the availability of precisely specified and, ideally, machine-verifiable requirements. MVCs with state-of-the-art performance rely on machine learning (ML) and training data, but largely lack such requirements. 

In this paper, we address the need for defining machine-verifiable reliability requirements for MVCs against transformations that simulate the full range of realistic and safety-critical changes in the environment.  
Using human performance as a baseline, we define reliability requirements as: `if the changes in an image do not affect a human’s decision, neither should they affect the MVC’s.’ 
To this end, we provide:
(1) a class of safety-related image transformations;
(2) reliability requirement classes to specify correctness-preservation and prediction-preservation for MVCs;  
(3) a method to instantiate machine-verifiable requirements from these requirements classes using human performance experiment data; 
(4) human performance experiment data for image recognition involving eight commonly used transformations, from about 2000 human participants; and
(5) a method for automatically checking whether an MVC satisfies our requirements.
Further, we show that our reliability requirements are feasible and reusable by evaluating our methods on 13 state-of-the-art pre-trained image classification models. Finally, we demonstrate that our approach detects reliability gaps in MVCs that other existing methods are unable to detect.

\end{abstract}



\maketitle

\input{introduction}

\input{overview}

\input{transformations}
\input{requirements}

\input{estimating}
\input{testing}

\input{evaluation_new}
\input{relatedwork}

\input{conclusion}

\section*{Acknowledgments}
The authors would like to thank the anonymous reviewers for their feedback and insightful comments.  We also thank all the MTurkers for participating in our experiments; Dr. Dimitrios Papadopoulos for providing an MTurk experiment implementation that formed the basis of our experiments; Professor Radu Craiu for his comments on the statistical methods; Dr. Nikita Dvornik, Nick Feng, Dr. Mona Rahimi, Valentina Manferrari, Dr. Ramy Shahin, Alexander Tough, and Dr. Shurui Zhou for helping improve this manuscript; and Valentina Manferrari for her assistance during an earlier version of the experiments. 

\bibliographystyle{plain}
\bibliography{main}
\end{document}

%% file: introduction.tex
\pdfoutput=1
\section{Introduction}
\label{introduction}
The use of Machine Vision Components (MVCs) in safety-critical systems, such as self-driving cars, creates major safety concerns, since undesired behaviors can lead to fatal accidents~\cite{zhang2018deeproad}.
For example, recently, Tesla self-driving cars misclassified emergency vehicles and caused multiple crashes~\cite{boudette-21,boudette-21-b}. 
Knowing how to analyze these components, provide safety assurance, and ensure their quality becomes a must for their usability in safety-critical domains.
Particularly, in systems that automate tasks normally performed by humans, such as driving, the vision task is performed by MVCs which represent a critical function for the overall system safety. 
However, vision tasks are difficult to specify; thus, they are usually performed using machine learning (ML)~\cite{Shuyuan-et-al-20}.
Defining requirements for ML is not trivial because the inability to specify clear requirements is the reason to use ML in the first place~\cite{picardi-20,Ishikawa-et-al-19,Vogelsang-19}.  Yet such requirements are necessary for verification and providing safety guarantees.
As a first step towards safe MVCs, one needs to define what it means for an MVC to be correct and then check its correctness prior to system deployment.

\begin{figure}
\centering
\scalebox{0.67}{
    \begin{tabular}{c|c|c|c}
        \midrule
        \multicolumn{2}{c|}{
             \begin{subfigure}{.16\textwidth}
             \includegraphics[width=\linewidth]{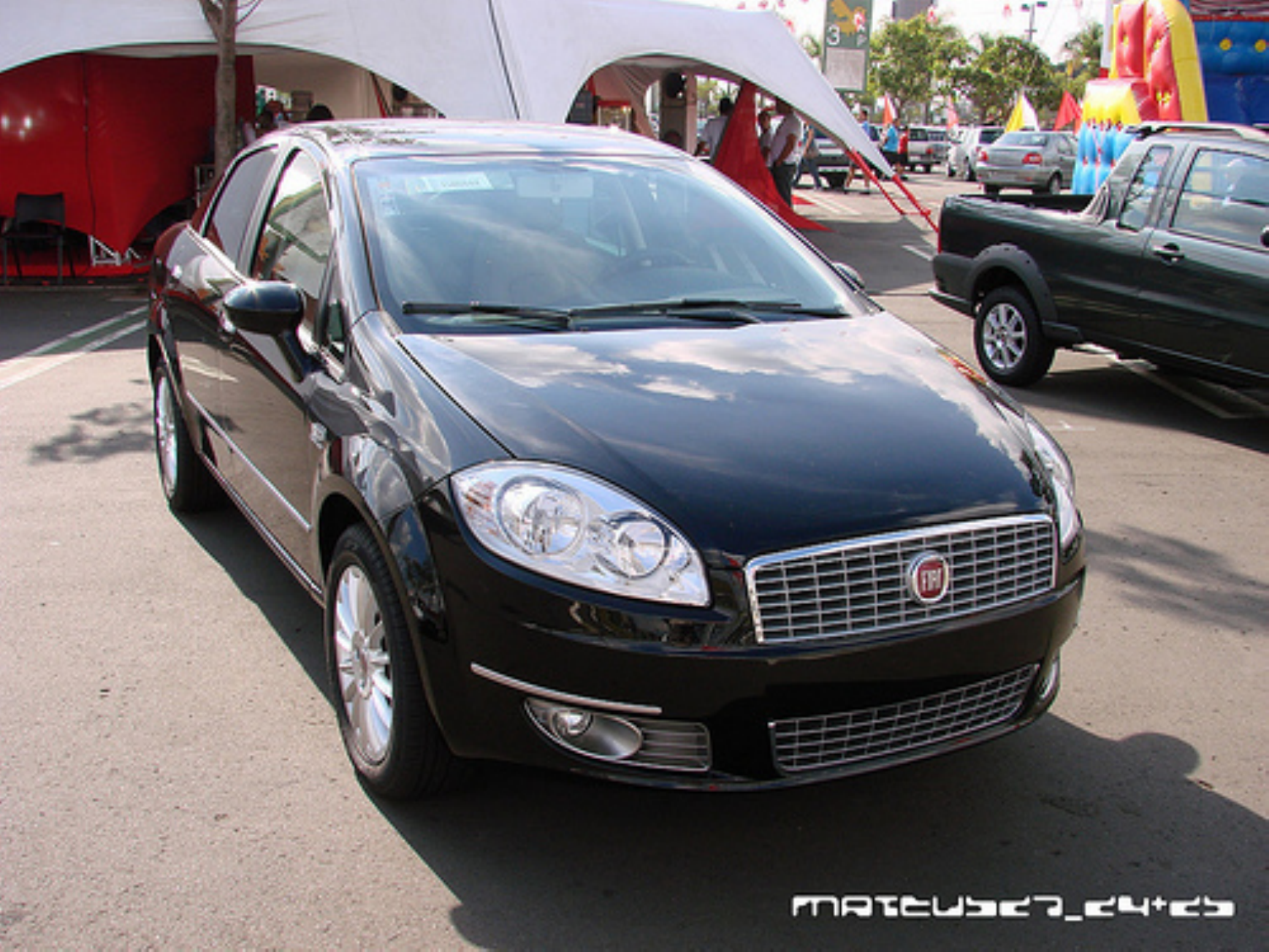}\caption{original image}\label{orig_0}
            \end{subfigure}
        } &
        \multicolumn{2}{c}{
             \begin{subfigure}{.16\textwidth}
             \includegraphics[width=\linewidth]{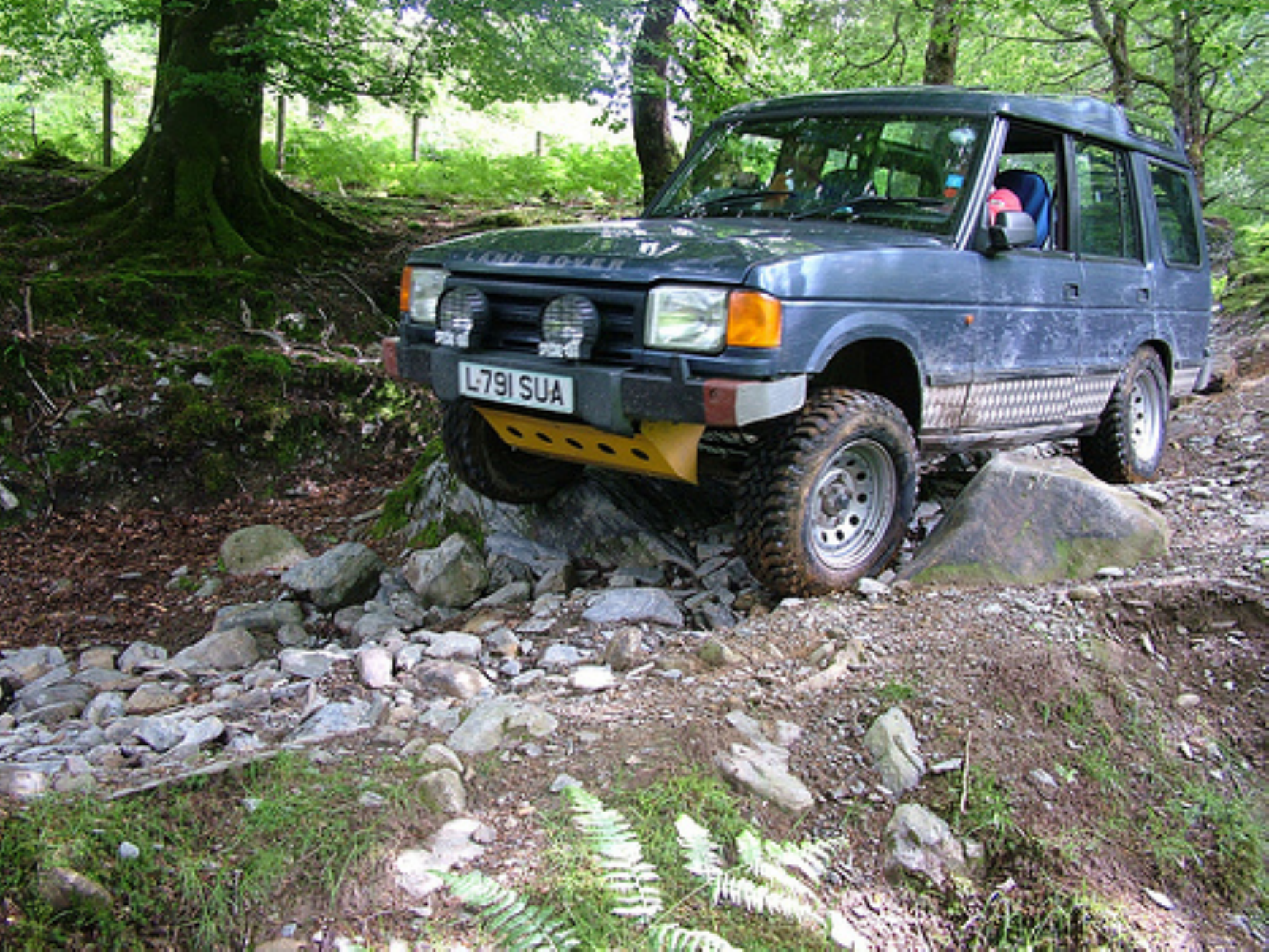}\caption{original image}\label{orig_1}
            \end{subfigure}
        } \\
        \multicolumn{2}{c|}{ \xmark \ \textit{not car (grille)} }& \multicolumn{2}{c}{ \cmark \ \textit{car (jeep)}} \\
        \midrule
         \begin{subfigure}{.16\textwidth}
            \includegraphics[width=\linewidth]{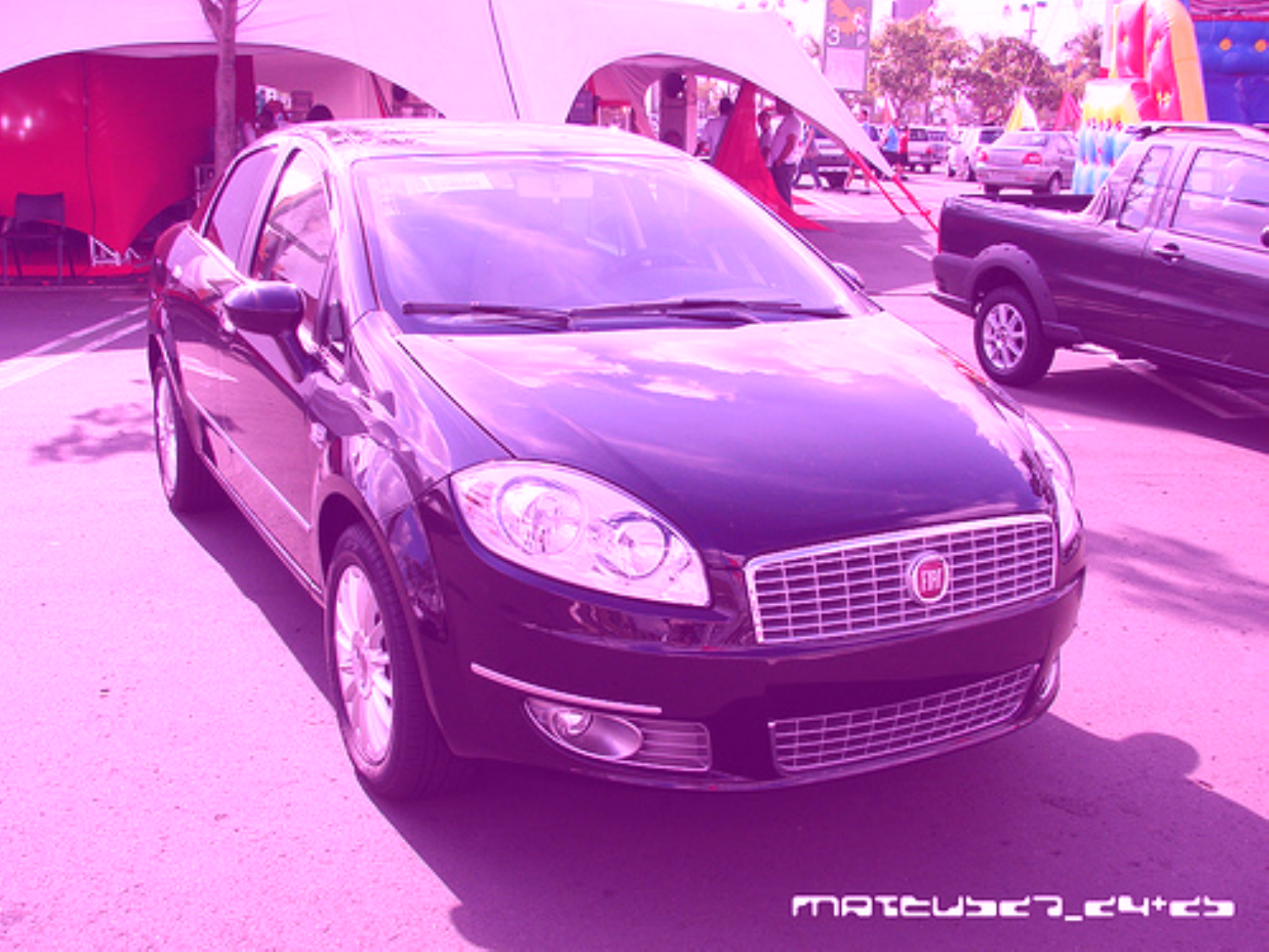}\caption{RGB}\label{rgb}
         \end{subfigure} &
         \begin{subfigure}{.16\textwidth}
            \includegraphics[width=\linewidth]{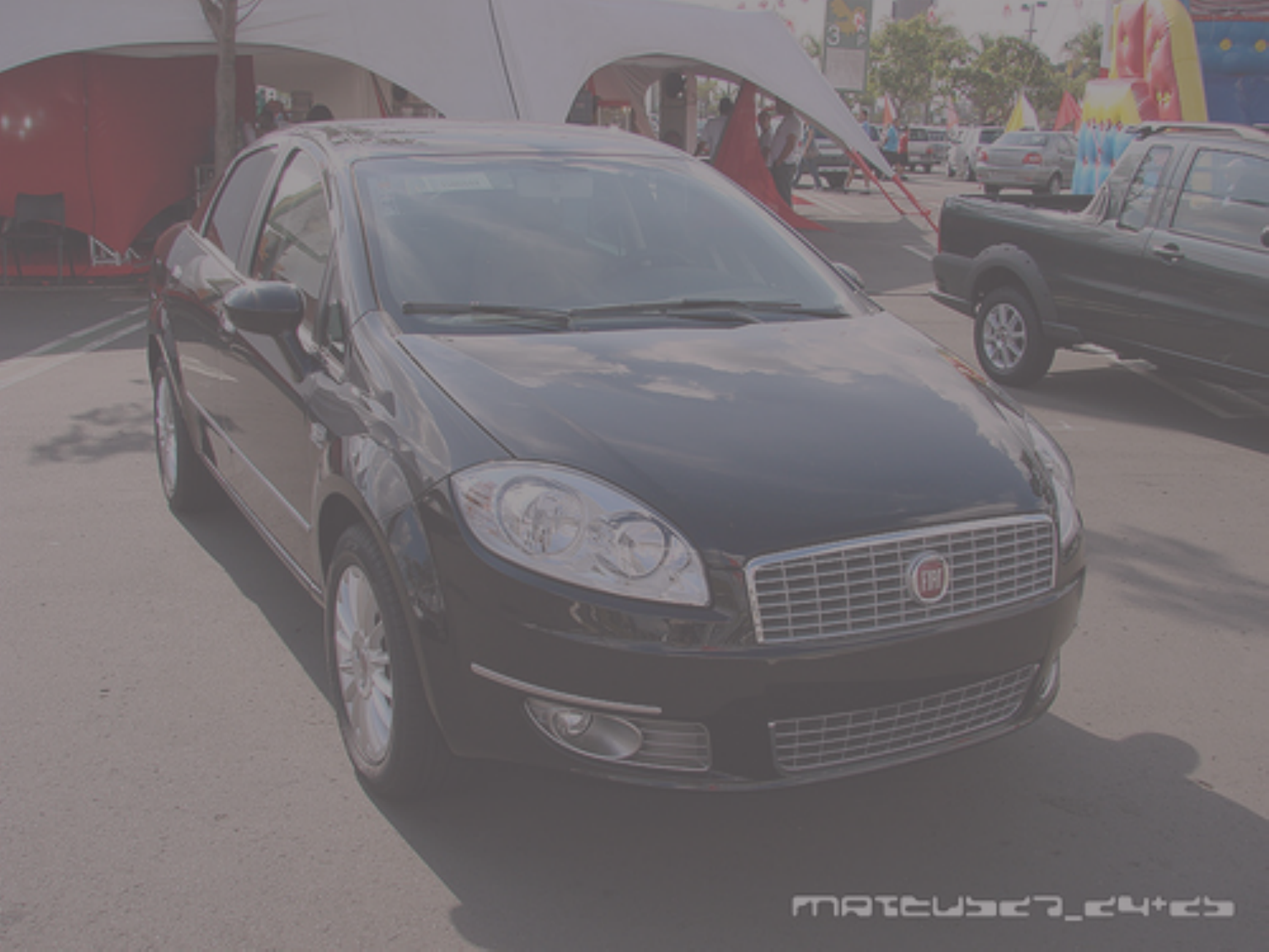}\caption{contrast}\label{contrast}
         \end{subfigure}  &
         \begin{subfigure}{.16\textwidth}
            \includegraphics[width=\linewidth]{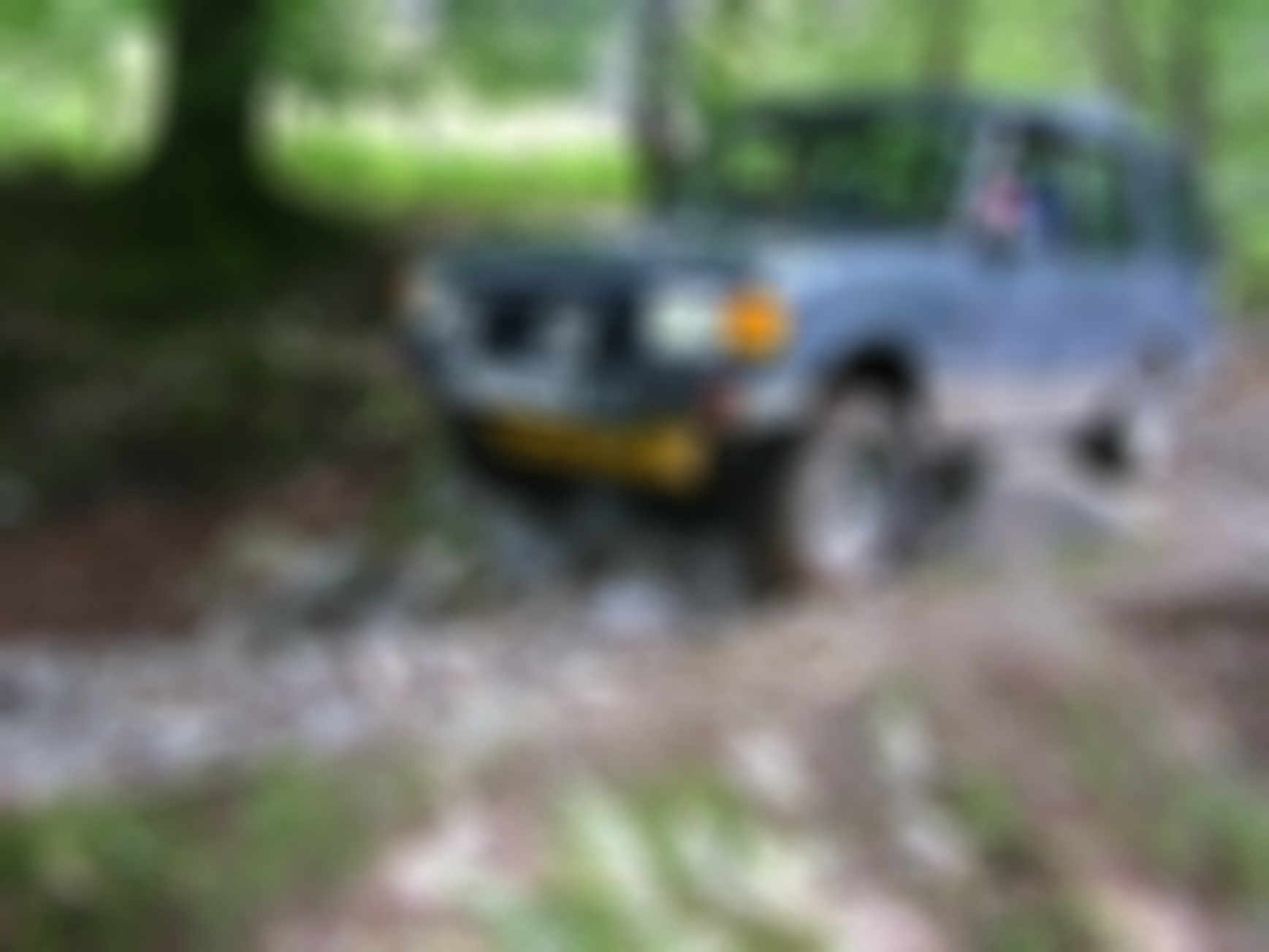}\caption{defocus blur}\label{defocus_blur}
         \end{subfigure} &
         \begin{subfigure}{.16\textwidth}
            \includegraphics[width=\linewidth]{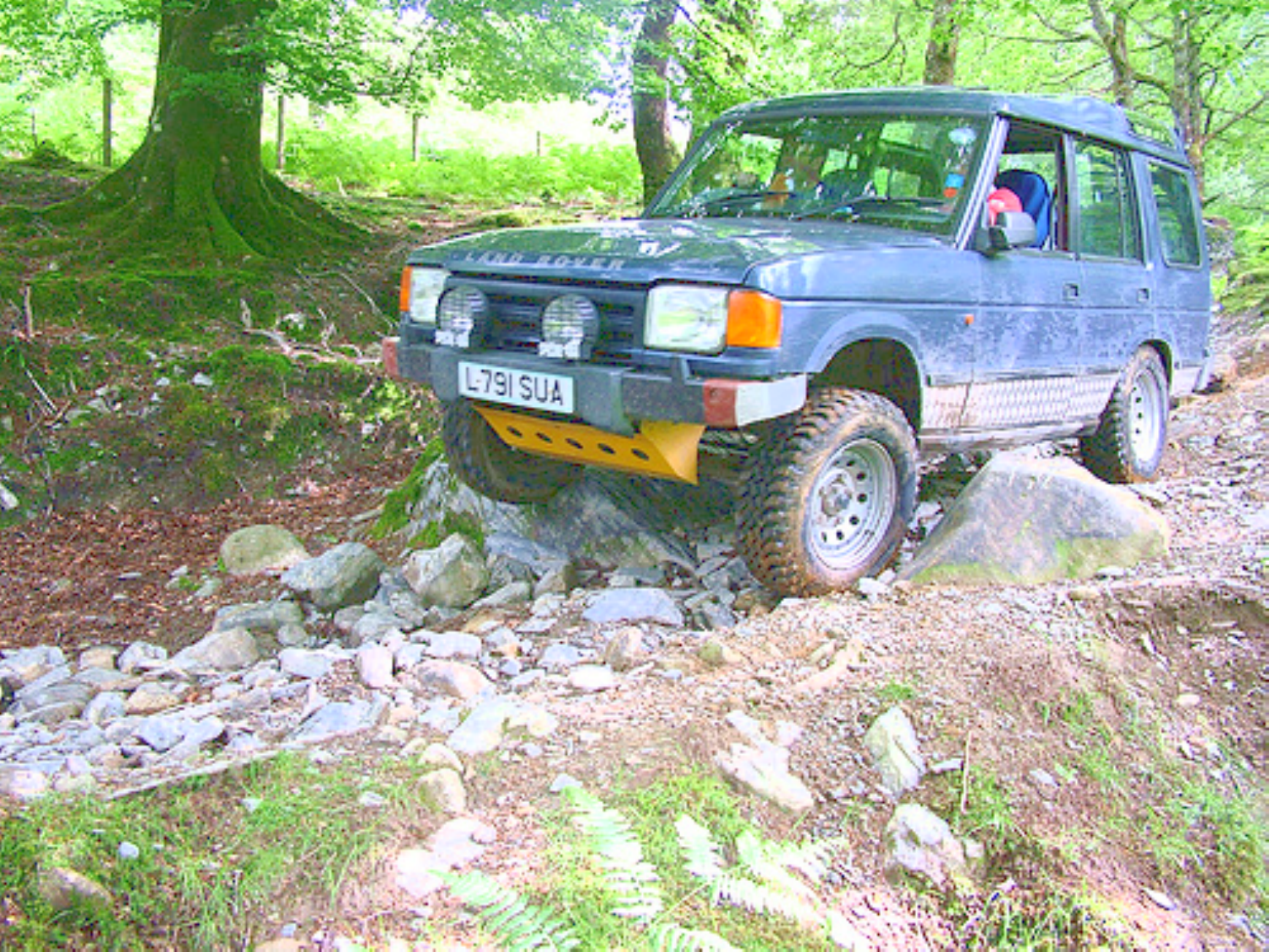}\caption{brightness}\label{brightness}
         \end{subfigure}   \\
          \cmark \ \textit{car (limousine)} & \xmark \ \textit{not car (grille)} & \xmark \ \textit{not car (EU salamander)} & \cmark \ \textit{car (jeep)}  \\
         \cmidrule{1-4}
         \begin{subfigure}{.16\textwidth}
            \includegraphics[width=\linewidth]{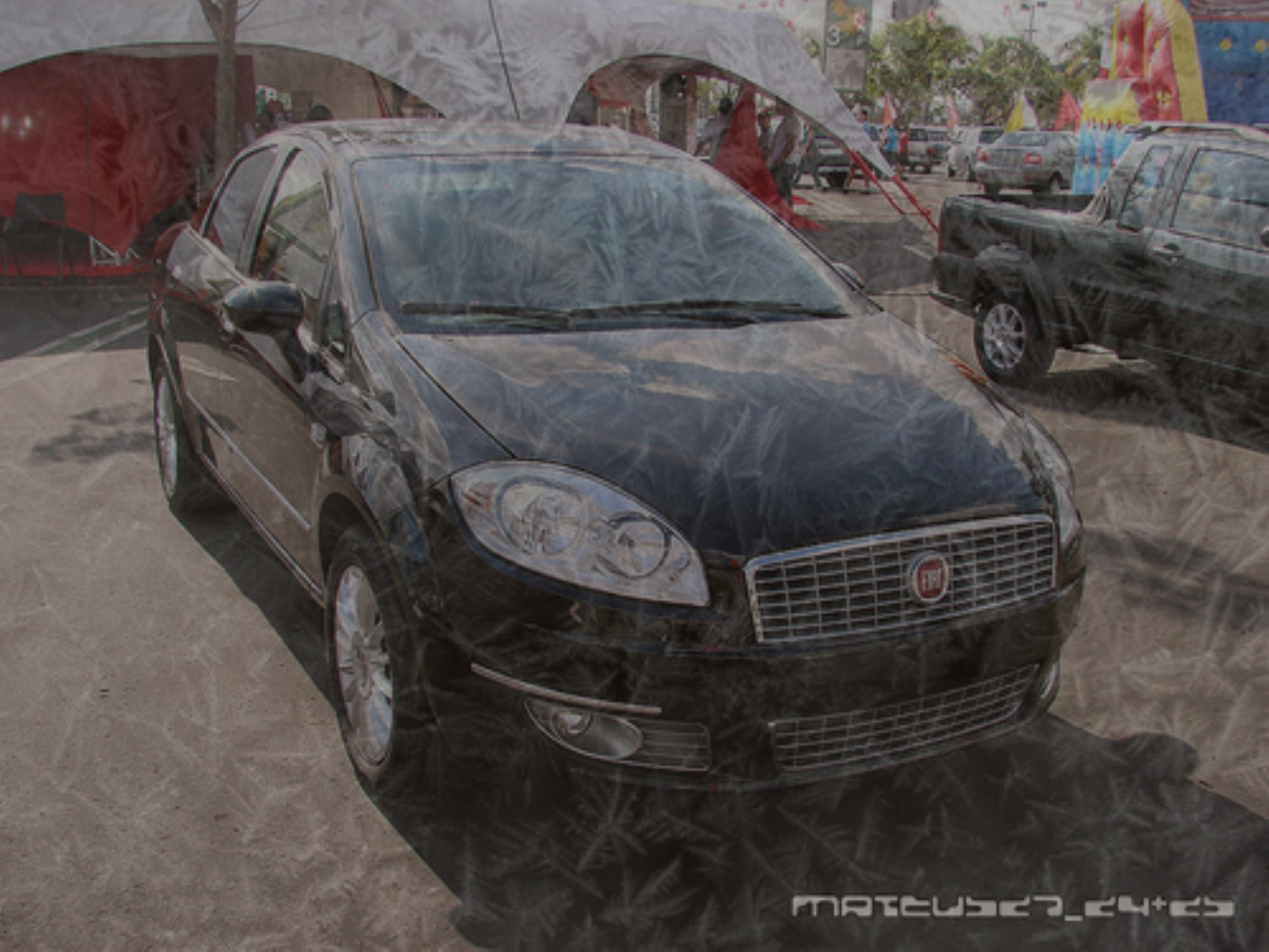}\caption{frost}\label{frost}
         \end{subfigure} &
         \begin{subfigure}{.16\textwidth}
         \includegraphics[width=\linewidth]{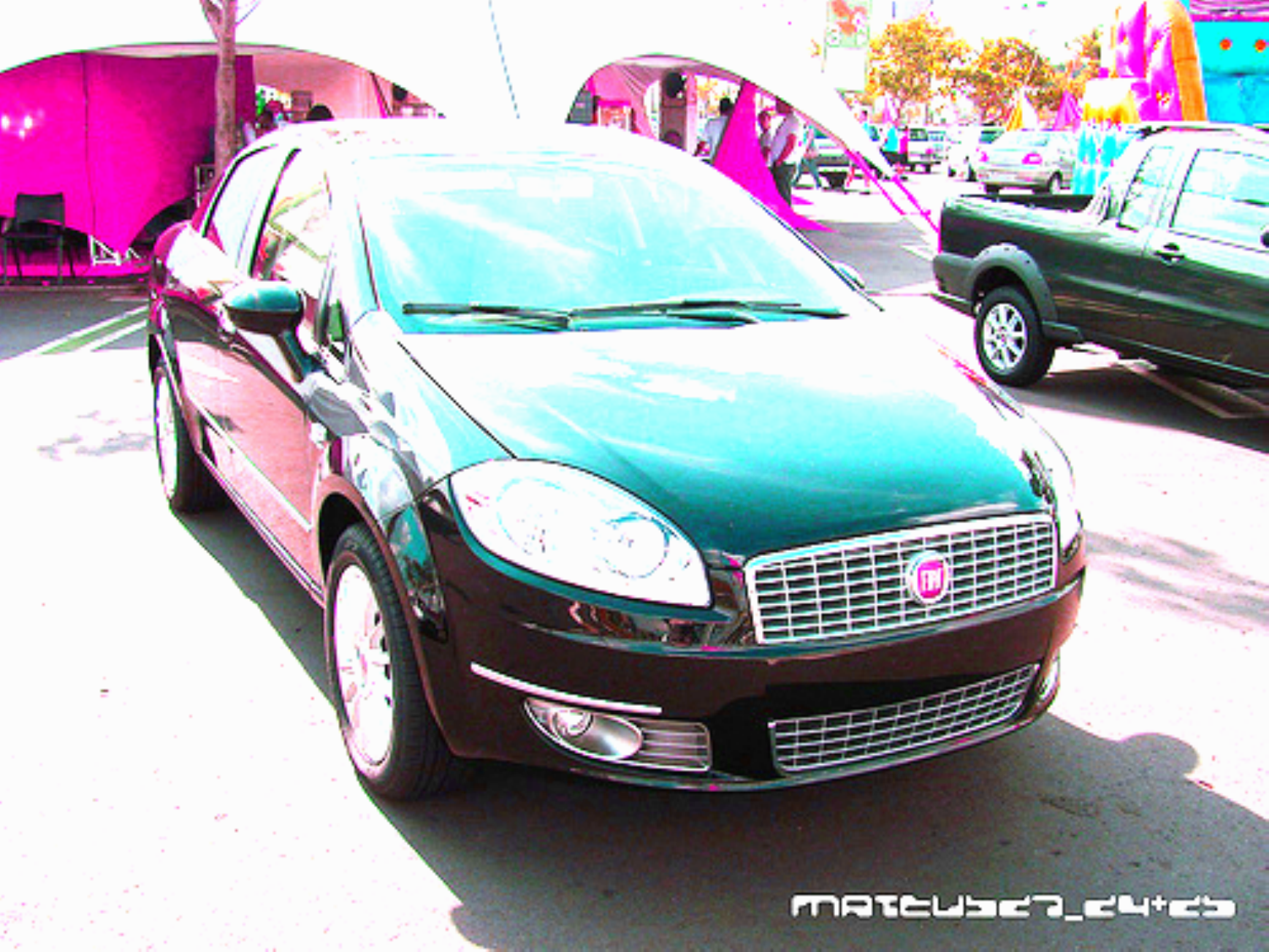}\caption{color jitter}\label{color_jiitter}
         \end{subfigure} & 
         \begin{subfigure}{.16\textwidth}
            \includegraphics[width=\linewidth]{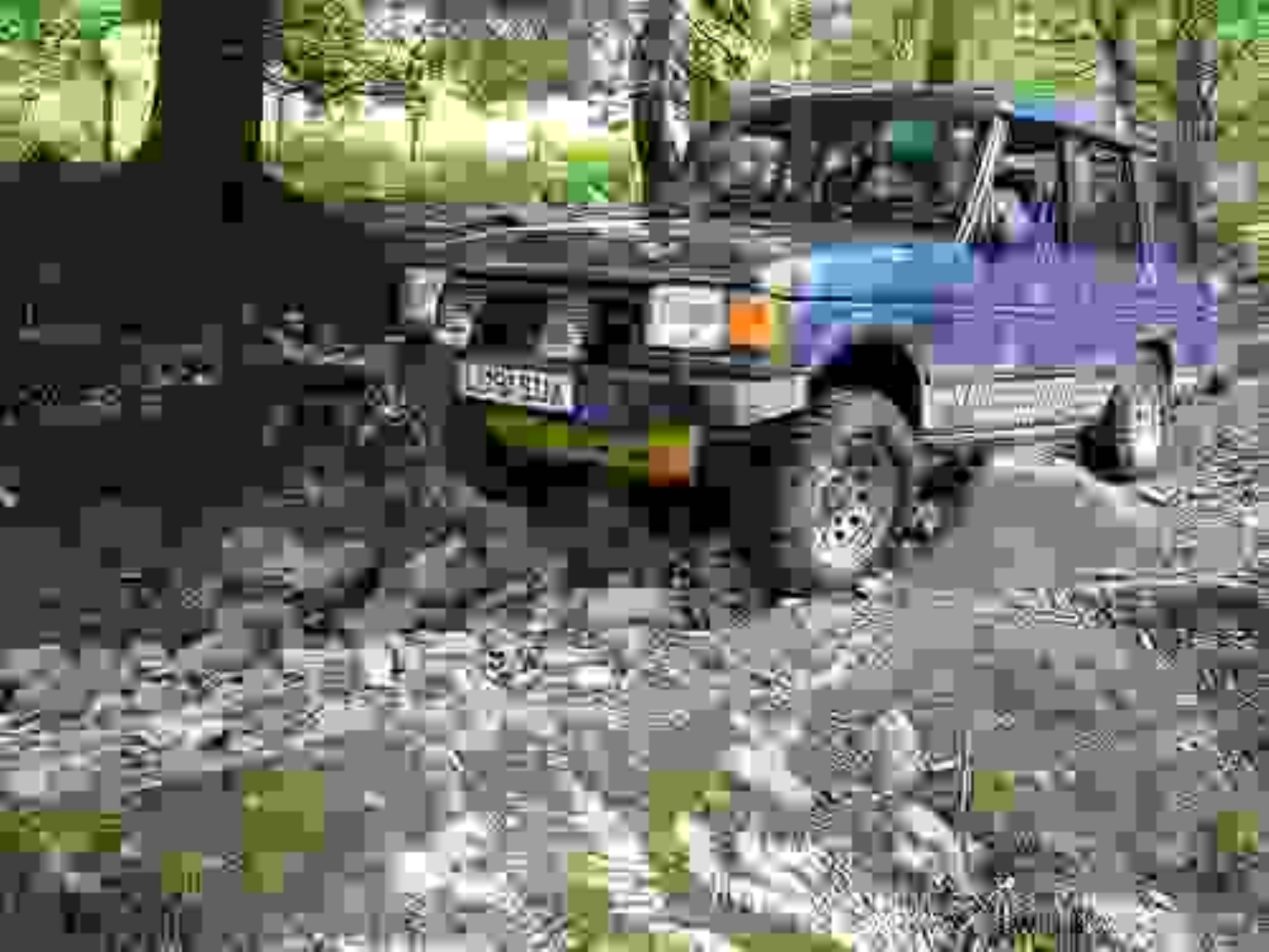}\caption{jpeg compression}\label{jpeg}
         \end{subfigure} &
         \begin{subfigure}{.16\textwidth}
            \includegraphics[width=\linewidth]{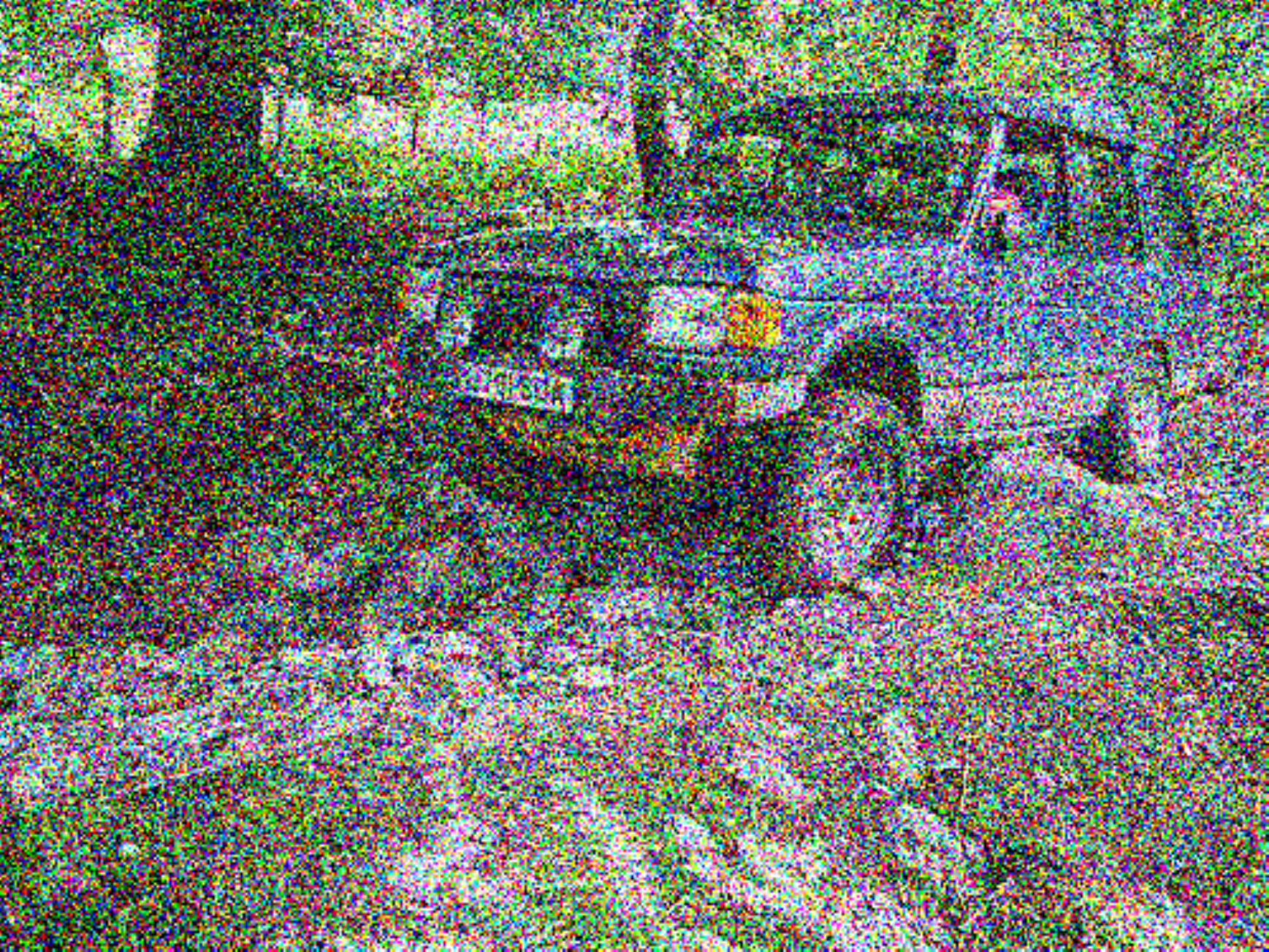}\caption{Gaussian noise}\label{gaussian_noise}
         \end{subfigure}\\
         \cmark \ \textit{car (minivan)} & \cmark \ \textit{car (minivan)} & \cmark \ \textit{car (jeep)} & \xmark \ \textit{not car (shovel)} \\
         \bottomrule
    \end{tabular}
}
\caption{\small Image recognition on original and transformed images.
    The top row displays original images containing cars from the ILSVRC'12 dataset~\cite{ILSVRC2012}. The transformation applied to the image is specified under the image. Images from (c) to (j) present all the safety-related transformations considered in this paper. 
    The classification result of a state-of-the-art MVC ResNet50~\cite{Resnet50} is shown in brackets in italics under each image. 
    We further specify whether the predicted category is concidered as a car (\cmark), or not (\xmark), based on the ILSVRC'12 class hierarchy. Transformations are implemented by Albumentations~\cite{albumentation} and Imagenet-c~\cite{hendrycks2019robustness}.
    }\label{fig:motivating}
    \vspace{-0.1in}
\end{figure}

In this paper, we focus on one aspect of correctness: \emph{reliability}, which measures the ability of a system or component to perform its required functions in a specified environment~\cite{IEEE_software_term}, as it enables ensuring the quality of the deployed system.
We are specifically interested in whether the performance of an MVC remains reliably unaffected by image transformations that commonly occur in real-world scenarios.
This question has been studied in SE and ML literature as \emph{model robustness}, including testing~\cite{xie2011metatesting} and verification~\cite{huang2017verif} techniques.
Yet, given the lack of detailed reliability requirements, these approaches are limited to checking the models within a
\emph{small} neighbourhood of the original input image, i.e., by applying perturbations that are almost imperceptible to humans. 
While considering only the small perturbations allows for requirement analysis of model reliability~\cite{AIRE2020}, its applicability is limited in the real-world scenarios, with a much broader range of possible changes.
For example, consider the problem of recognizing cars in images -- see a few examples in Fig.~\ref{fig:motivating}.  We are interested in being able to recognize cars under such transformations as frost  (see Fig.~\ref{frost}) and different brightness levels (Fig.~\ref{brightness_small} and Fig.~\ref{brightness_big}).

The range of transformation magnitudes in images in Fig.~\ref{fig:motivating} is not considered small or imperceptible. While humans have no problem recognizing cars in these images, the state-of-the-art image classification model ResNet50~\cite{Resnet50} failed to do so on the examples in  Figs.~\ref{contrast}, \ref{defocus_blur} and \ref{gaussian_noise}.
Since MVCs are used in systems that automate tasks normally performed by humans, MVCs, like ResNet50, are \emph{at the minimum} expected to consistently classify objects across range of changes that do not affect human perception.
Thus, we seek a method to establish  human performance as a reference for defining reliability, and an automated method to check MVC against a justified range of changes that do not affect human performance.

In this paper, we formally define two classes of machine-verifiable reliability requirements for MVCs: \emph{correctness-preservation} and \emph{prediction-preservation}. 
For both requirements classes, the range of image changes we consider (i.e., the human-tolerated range), is a parameter estimated using experiments with human participants. 
Intuitively, within the human-tolerated range of changes, \emph{correctness-preservation} requires that the MVC's predictions after changes in images should be correct, and \emph{prediction-preservation} requires that the predictions on original images and on images that underwent transformations should be the same.  Specifically, this paper makes the following contributions:
(1) We identify a class of safety-related image transformations;  
(2) We provide a formal specification of two classes of input-output reliability requirements for MVCs, with parameters representing human performance;
(3) We present a method to instantiate our requirements classes into machine-verifiable requirements. This method estimates ranges of changes to images that do not affect human vision using results of experiments with human participants;
(4) We provide human experiment performance data for image recognition;
(5) We provide an automated method for checking  MVCs against our machine-verifiable requirements.

While our criteria are defined for any computer-vision task (including object detection and semantic segmentation), in this paper, we demonstrate the feasibility of our approach on the image classification task.
We show that our approach captures reliability gaps that existing methods are unable to detect using 13 state-of-the-art pre-trained image classification models on two image classification datasets (Imagenet~\cite{ILSVRC2012} and CIFAR-10~\cite{krizhevsky-09}).

\vskip 0.05in
\noindent{\bf Significance:}  
To the best of our knowledge, we are the first to define reliability requirements for MVCs using a human-justified range of changes over realistic safety-related transformations.    Our requirements and the method for checking their satisfaction can be reused by software engineers for analyzing system reliability of MVCs before deployment.

The rest of the paper is organized as follows: Sec.~\ref{sec:overview} gives an overview of our approach for creating and checking our reliability requirements. 
Sec.~\ref{transformations} presents the safety-related image transformations and a generic metric for measuring changes in images.
Sec.~\ref{requirements} presents a  formal specification of our reliability requirement classes.
Sec.~\ref{estimate}  presents our experiment for measuring human recognition performance with human participants, and demonstrates an automated approach for estimating parameters of the requirements, using data from this experiment. 
Sec.~\ref{testing} introduces an automated method for checking MVC's against our reliability requirements. 
We evaluate our approach in Sec.~\ref{sec:evaluation}.  
Sec.~\ref{sec:rw} compares our work with related approaches and we conclude in Sec.~\ref{conclusion}.

%% file: overview.tex
\pdfoutput=1
\section{Approach Overview}
\label{sec:overview}
\input{process_figure}
Fig.~\ref{fig:flowchart} gives an overview of our approach.  Given (i) a vision task for the MVC, (ii) a safety-related transformation and (iii) experimental data for estimating the ranges of visual changes that do not affect human performance, we provide a process for instantiating machine-verifiable reliability requirements for MVC (requirement instantiation) and a process for checking whether an MVC satisfies these instantiated requirements (requirement checking).

The vision task and the transformation need to be selected based on the application of the MVC.
To help with the selection of transformations that represent changes likely to happen in the operating environment, we identified a class of safety-related image transformations that represent potentially risky input modifications in real world situations. 
For example, frost shown in Fig.~\ref{fig:motivating} is a safety-related transformation because it can reduce lighting in the scene which, in turn,  can cause machine vision errors.
Note that since transformations have different parameter domains and can have different visual effects on different images to humans, we defined a generic metric called a \emph{visual change} and denoted by $\Delta_v$, which decouples the perceptible visual change to the image from the transformation parameters and thus allows stating the reliability requirements on the MVC more abstractly.

\vspace{0.05in}
\noindent
\textbf{Requirement instantiation:} 
This automated step enables users to instantiate the reliability requirements for the vision task with human tolerated range of visual changes for each selected transformation.
The human-tolerated range is the requirement parameter that describes the range of changes from a transformation that should not affect the MVC's behavior. 
This requirement parameter is measured with $\Delta_v$ and estimated using results from experiments with human participants.
The output of this step is a set of machine-verifiable requirements. The resulting correctness-preservation [resp. prediction-preservation] requirement states: 
for a vision task and a transformation, if the changes in the images are within the estimated human tolerated range, then an MVC should preserve the correctness [resp. prediction] after applying the changes to its input from before. 

For example, for the transformation adding frost artificially, our resulting requirements are as follows:
\begin{itemize}
    \item the recognition accuracy of an MVC should not decrease if the visual change in the images is within the range $\Delta_v \leq 0.84$ (\emph{correctness-preservation}); and 
\item the percentage of labels the MVC can preserve after adding frost should not decrease if visual change in the images is within the range $\Delta_v \leq 0.91$ (\emph{prediction-preservation}).
\end{itemize}

Note that our requirements do not depend on the state-of-art ML techniques since they treat the MVC as a black box.

\vspace{0.05in}
\noindent
\textbf{Requirement checking:} 
This automated method checks whether an MVC satisfies the instantiated reliability requirements.
Given a set with original images, our process generates test cases (step~\RN{2}.a) by transforming the original images within the range specified in the instantiated requirements, runs the tests on the model (step~\RN{2}.b), and checks whether the MVC satisfies our requirements (step~\RN{2}.c).

To summarize, our proposed approach can be used to automatically generate machine-verifiable reliability requirements for a vision task and a list of transformations, given human experiment results; and then automatically evaluate whether an MVC satisfies these requirements.
In the above example, the requirement checking method will generate a set of test case images within the ranges ($0.84$ and $0.91$) to check whether an MLC satisfies these requirements. 

An implementation of our method is available online.\footnote{\label{gitlink_note} See \url{https://carolineeeeeee.github.io/automating_requirements} 
for implementation, more results and information.} 
For the purpose of demonstration and evaluation, we conducted image classification experiments with 2000 human participants for the vision task of recognizing car images for 8 transformations: RGB, contrast, defocus blur, brightness, frost, color jitter, jpeg compression, and Gaussian noise (see images in Fig.~\ref{fig:motivating}(c)-(j)).  
In the rest of the paper, we describe the technical details of each step of Fig.~\ref{fig:flowchart} using this experiment.

%% file: process_figure.tex
\pdfoutput=1
\tikzset{
    rndoutputitt/.style={draw,rectangle,rounded corners,minimum width=2cm,minimum height=1.5cm,align=center,text width=2.4cm,fill=blue!20,font=\small\sffamily},
    rndoutputi/.style={draw,rectangle,rounded corners,minimum width=2cm,minimum height=1.5cm,align=center,text width=3.2cm,fill=blue!20,font=\small\sffamily},
    rndoutputis/.style={draw,rectangle,rounded corners,minimum width=2cm,minimum height=0.9cm,align=center,text width=3.2cm,fill=blue!20,font=\small\sffamily},
    rndoutputib/.style={draw,rectangle,rounded corners,minimum width=2cm,minimum height=0.9cm,align=center,text width=3.5cm,fill=blue!20,font=\small\sffamily},
    rndoutputl/.style={draw,rectangle,rounded corners,minimum width=0.65cm,minimum height=0.6cm,align=center,text width=0.75cm,fill=blue!20,font=\sffamily},
    rndoutput/.style={draw,rectangle,rounded corners,minimum width=2cm,minimum height=1.5cm,align=center,text width=2cm,fill=blue!20,font=\small\sffamily},
    extact/.style={draw,dashed,rectangle,rounded corners,minimum width=2cm,minimum height=1.5cm,align=center,text width=3.4cm,fill=gray!50,font=\small\sffamily},
    recnorm/.style={draw,minimum width=0.6cm,minimum height=0.65cm,align=center,text width=0.65cm,fill=lightgray!50,font=\small\sffamily},
    recnormw/.style={draw,minimum width=0.6cm,dashed,minimum height=0.65cm,align=center,text width=0.65cm,fill=white,font=\sffamily},
    reclong/.style= {rounded corners=0.5cm,draw,minimum width=2cm,minimum height=1.5cm,align=center,text width=4cm,fill=gray!35,font=\sffamily},
    prog/.style={draw,dashed,tape,tape bend top=none,align=center,text width=2cm,fill=green!20,font=\small\sffamily},
    progl/.style={draw,tape,tape bend top=none,align=center,text width=0.7cm,fill=green!20,font=\small\sffamily},
    artifact/.style={draw,trapezium,trapezium left angle=75,trapezium right angle=105,text width=2.5cm,fill=yellow!35,align=center,font=\small\sffamily},
    artifactm/.style={draw,trapezium,trapezium left angle=83,trapezium right angle=98,text width=2.7cm,fill=yellow!35,align=center,font=\small\sffamily},
    artifacti/.style={draw,trapezium,trapezium left angle=83,trapezium right angle=98,text width=2.3cm,fill=yellow!35,align=center,font=\small\sffamily},
    artifactitt/.style={draw,trapezium,trapezium left angle=83,trapezium right angle=98,text width=2cm,fill=yellow!35,align=center,font=\small\sffamily},
    artifacto/.style={draw,trapezium,trapezium left angle=79,trapezium right angle=100,text width=1.5cm,fill=yellow!35,align=center,font=\small\sffamily},
    artifactoi/.style={draw,trapezium,trapezium left angle=79,trapezium right angle=100,text width=1.5cm,fill=lightgray!50,align=center,font=\small\sffamily},
    artifacts/.style={draw,dashed,trapezium,trapezium left angle=79,trapezium right angle=100,text width=3cm,fill=gray!50,align=center,font=\small\sffamily},
    artifactl/.style={draw,trapezium,minimum height=0.35cm,trapezium left angle=75,trapezium right angle=105,text width=0.4cm,fill=yellow!35,align=center,font=\sffamily},
    artifactvt/.style={draw,trapezium,trapezium left angle=83,trapezium right angle=98,text width=1cm,fill=yellow!35,align=center,font=\small\sffamily}
}

\begin{figure}[t]
    \centering
\scalebox{.60}{
\begin{tikzpicture}[x=2.25cm,y=0.9cm]

\draw (-0.87,-4) rectangle (5.4,-2.75);
\draw (-0.87,-2.75) rectangle (2.55,6.5);
\draw (2.55,-2.75) rectangle (5.4,6.5);

\node[rotate=0,font=\sffamily] at (-0.5,-3.2) {\shortstack{\underline{Legend:}}};
\node[rotate=0,font=\sffamily] at (0.7,6) {\textbf{\shortstack{\small \RN{1}.Requirement instantiation}}};
\node[rotate=0,font=\sffamily] at (4,6) {\textbf{\shortstack{\small \RN{2}.Requirement checking}}};

\node[rotate=0,font=\small\sffamily] at (0.6,-3.2) {\shortstack{:~Step}};
\node[rotate=0,font=\small\sffamily] at (1.8,-3.2) {\shortstack{:~Artifact}};

\node[artifactvt] (img_safety) at (-0.45,1.6) {Vision task};
\node[artifactvt] (transf) at (0.3, 1.6) {Safety-related transformation};
\node[artifactitt] (h_data) at (0, -0.6) {Experiment data of human performance};
\node[rndoutputib] (estimate) at (1.6,-0.7)  {\RN{1}.a. Estimate human tolerated range of changes with experiment results for the vision task and for each transformation};
\node[artifact, above = 12pt of estimate] (range) {Parameters for requirement classes};
\node[rndoutputib, above = 12pt of range] (spec) {\RN{1}.b. Instantiate the reliability requirements};
\node[artifactm] (output) at (1.6,4.4) {Machine-verifiable requirements};
\node[rndoutputi] (test_gen) at (3.45,4.6)  {\RN{2}.a. Generate tests for checking the satisfaction of requirements};
\node[artifacto, right = 16pt of test_gen] (img_orig_set) {Images sets};
\node[artifact, below = 12pt of test_gen] (img_set) {Set of test images (original and transformed)};
\node[rndoutputi, below = 18 pt of img_set] (check) {\RN{2}.c Estimate confidence of requirements satisfaction (reliability distance)};
\node[rndoutput, right = 16pt of img_set] (ml_model) {\RN{2}.b. Run the tests on the MVC under validation};
\node[artifacto, right = 12pt of check] (prediction) {MVC prediction results};
\node[artifact, below = 12pt of check] (confidence) {Requirements satisfaction \\results};
\node[rndoutputl] (activ) at (0.2,-3.3) { };
\node[artifactl, right = 37pt of activ] (artifact) {~~~};

\begin{scope}[every path/.style={-latex}]
\draw[dashed,->] (img_safety) -- (h_data);
\draw[dashed,->] (transf) -- (h_data);
\draw 
    (h_data) edge (estimate)
    (estimate) edge (range)
    (range) edge (spec)
    (spec) edge (output)
    (output) edge (test_gen)
    (img_orig_set) edge (test_gen)
    (test_gen) edge (img_set)
    (img_set) edge (ml_model)
    (prediction) edge (check)
    (ml_model) edge [bend left] (prediction)
    (check) edge (confidence);
\end{scope}
\end{tikzpicture}
}

\caption{\small A process for instantiating  two reliability requirements classes for MVCs (requirement instantiation) and a process for checking their satisfactions (requirement checking). }
    \label{fig:flowchart}
    \vspace{-0.18in}
\end{figure}
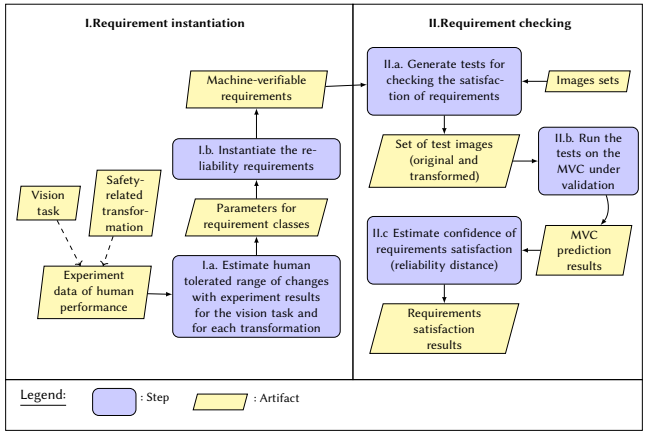

%% file: transformations.tex
\pdfoutput=1
\vspace{-0.05in}
\section{Visual changes in images}
\label{transformations}

In this section, we start with establishing the definition of the metric $\Delta_v$, which measures human visual changes in images caused by transformations.
Then, we identify a class of safety-related image transformations that are used to instantiate our \emph{correctness-preservation} and \emph{prediction-preservation} requirements (see Sec.~\ref{requirements}).

A key idea in our work is to define reliability requirements relative to $\Delta_v$ ranges rather than the transformation parameter ranges to be tolerated. This is important since each transformation may have one or more parameters, and each parameter may affect the transformed image to a different degree---also depending on the input image. For example, brightening an already bright image makes the objects harder to see; on the other hand, making a dark image brighter will have the opposite effect. 
Further, small changes to one parameter may cause small changes or large changes to the transformed image depending on the values of other parameters.
The visual change metric $\Delta_v$ allows us to abstract from these complexities of the transformation parameter space. Also, a simple image distance metric such as mean squared error, which is often used to define robustness, e.g., \cite{bastani2016measuring,cohen2019certified},
does not adequately reflect the human-perceived visual change in images~\cite{Wang-MSE}. Thus, we base  $\Delta_v$ on image quality assessment metrics.

\vskip 0.05in
\noindent
{\bf Background: Image Quality Assessment (IQA)}. 
IQA metrics are quantitative measures of human objective image quality~\cite{SSIM}. Given the original image, the IQA metrics automatically predict the perceived image quality by measuring the perceptual `distance' between the two images~\cite{VIF}. This `distance' is different from pixel distance and its calculation depends on the design of the IQA metric. VSNR (\emph{Visual-Signal-to-Noise-Ratio}~\cite{VSNR}) checks the visibility of the changes in images and returns infinity ($\infty$) if they  are not visible to humans~\cite{VSNR}. VIF (\emph{Visual Information Fidelity}~\cite{VIF}) measures the information fidelity by analyzing the statistics of the natural scenes in the images. VIF returns a value between 0 and 1 if the changes degrade perceived image quality, with 1 indicating the perfect quality compared to the original image; and it returns a value  $>1$ if the changes enhances image quality~\cite{VIF}. VIF is empirically shown to be the closest 
to human opinions when compared to all other IQA metrics~\cite{Sheikh2006} and VSNR has been shown to be effective to detect non-visible changes~\cite{AIRE2020}.
VIF is applicable to transformations that can be described locally by a combination of signal attenuation and additive Gaussian noise in the sub-bands in the wavelet domain~\cite{VIF}.

\vskip 0.05in
\noindent
{\bf Measuring visual change in images.}
We now use IQA metrics to define a generic metric $\Delta_v$.  Our definition of $\Delta_v$  shares the same applicability characteristics as VNSR and VIF.  For transformations that satisfy this characteristic (e.g., noise, blur, brightness and contrast changes, color change, etc.), the definition is as follows:
\begin{mytheo}{Visual change $\Delta_v$}{}
Let an image $x$, an applicable transformation $T_X$ with a parameter domain $C$ and a parameter $c\in C$, s.t. $x' = T_X(x,c)$ be given.
$\Delta_v(x,x')$ is a function defined as follows:
    {\small  \[ \begin{cases} 
      0  & $If VSNR$(x, x')= \infty \\
      &$ or VIF$(x,x')>1\\
       1-$VIF$(x, x')& $Otherwise  $
   \end{cases}
\]}
\label{def:visual_degrade}
\end{mytheo}
Basing $\Delta_v$ on IQA metrics means that it provides a generalized quantitative measure for visual changes in the images that is independent of particular images and transformations. 
We split this definition into two cases. The first corresponds to changes imperceptible to humans 
(when VSNR$(x, x')= \infty$) and changes that enhance the visual quality (when VIF$(x,x')>1$). 
In this case, $\Delta_v=0$ because such changes do not impact human recognition of the images negatively.
The other case deals with visible changes that degrade visual quality.  Since
VIF returns 1 for perfect quality compared to the original image, the degradation is one minus the image quality score.
For example, 
the visual change of the example in Fig.~\ref{brightness} compared to its original image in Fig.~\ref{orig_1} is $0.507$.   The visual change of the example in Fig.~\ref{defocus_blur} compared to the one in Fig.~\ref{orig_1} is $0.985$. This suggests that the transformation in Fig.~\ref{defocus_blur} causes more change visually than the one in Fig.~\ref{brightness}.

\vskip 0.05in
\noindent
{\bf Safety-related image transformations.}
We say that a transformation is \emph{safety-related} if it can lead to a hazard in a real-world machine-vision application scenario. 
To assess this in a systematic manner, we utilize the CV-HAZOP checklist~\cite{CV_HAZOP}. This checklist comprehensively identifies the potentially hazardous impacts of different modes of interference in the computer vision (CV) process, which is comprised of light sources, transmission medium, object, observer, and algorithm. 
A transformation that can produce such impacts is considered safety-related.
For example, contrast adjustment, defocus blur, and added Gaussian noise shown in Fig.~\ref{fig:motivating} are safety-related transformations because they can reduce lighting in the scene, cause blurring, and add noise in the images, which can cause machine vision errors and subsequent system failures.
Since the scope of CV-HAZOP is broader than the image transformation assessment task, we remove non-image-related hazard scenarios entries from the checklist. 
In particular, entries related to \textit{Algorithm} in the vision process are not relevant because they modify the CV algorithm and not the images. Entries concerned with the \textit{Number} of \textit{Observers} are also not relevant since they focus on the interaction between the observers and cannot be represented by single image transformations. Finally, image transformations cannot make temporal changes; therefore, entries which deal with time are not relevant either.

To determine whether a given image transformation belongs to our safety-related class, one should first identify the location in the vision process to which the transformation corresponds; then the property of the process location that the transformation is affecting (CV-HAZOP parameters); and finally, how the transformation is changing the property (CV-HAZOP guide words). For example, defocus blur is changing the focus of the observer (CV-HAZOP entry No.1018), i.e., camera, and therefore belongs in our class. Supplementary material$^{\ref{gitlink_note}}$ includes the full list of CV-HAZOP safety-related entries (954 entries chosen from the overall 1470).

In this paper, we consider transformations provided by the state-of-the-art library Albumentations~\cite{albumentation} and the ML robustness benchmark Imagenet-c~\cite{hendrycks2019robustness}, which consist of 50 unique transformations.  Of these, 45  are safety-related.  We further remove those transformations that cannot produce a continuous range of transformed images, yielding 31 transformations---see supplementary material$^{\ref{gitlink_note}}$ for the full list.
Since multiple transformations can correspond to a single CV-HAZOP entry, we only  instantiate our approach on one transformation per CV-HAZOP entry, resulting in the eight transformations illustrated in Fig.~\ref{fig:motivating}c-j: \emph{RGB}, \emph{contrast}, \emph{defocus blur}, \emph{brightness}, \emph{frost}, \emph{color jitter}, \emph{jpeg compression} and \emph{Gaussian noise addition}.

%% file: requirements.tex
\pdfoutput=1
\vspace{-0.11in}
\section{Specification of Reliability Requirements Classes}
\label{requirements}
In this section, we provide a formal specification of our two reliability requirements classes: \emph{correctness-preservation} and \emph{prediction-preservation}.

Let us assume that we are given an MVC $f$, a distribution of input images $P_X$, a ground-truth labeling function $f^*$, a transformation $T_X$ with parameter domain $C$ and parameter distribution $P_C$. Our requirements use the joint distribution of
pairs of original and transformed images, defined as $P_{T_X}(x, x') = P_X(x)\sum_{c\in C, x'=T_X(x, c)} P_C(c)$.

We first introduce our \emph{correctness-preservation} reliability requirement class. It assumes a performance measure $m(f,f^*,P_X)$, which is typically a measure of similarity between the output of $f$ and $f^*$ given that the input $x\sim P_X$. Note that $m$ should be adequate for the vision task, such as classification accuracy for image classification, intersection over union (IoU) for image segmentation, and average precision for object detection. We define the marginal distribution of transformed images with changes within the human tolerated-range $\Delta_v \leq t_c$ as $P_{T_{X, t_c}}(x')= \sum_{x\in X} P_{T_X}(x, x' |  \Delta_v(x, x') \leq t_c)$.

\begin{figure}[t]
\scalebox{0.9}{
    \begin{tabular}{c|c|c}
     \toprule
            \begin{subfigure}{.16\textwidth}
             \includegraphics[width=\linewidth]{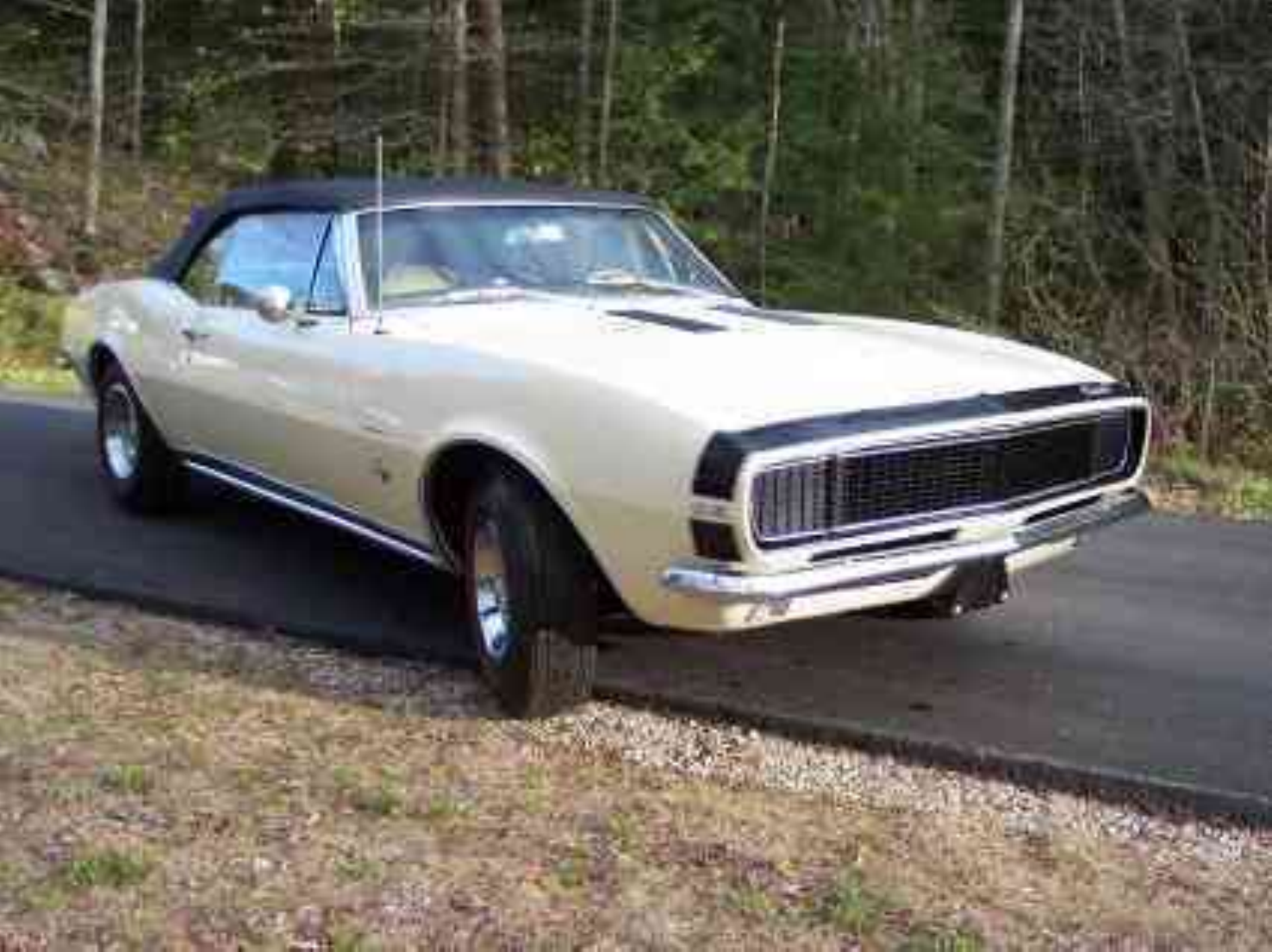}\caption{original image}\label{orig_2}
            \end{subfigure} &
            \begin{subfigure}{.16\textwidth}
                \includegraphics[width=\linewidth]{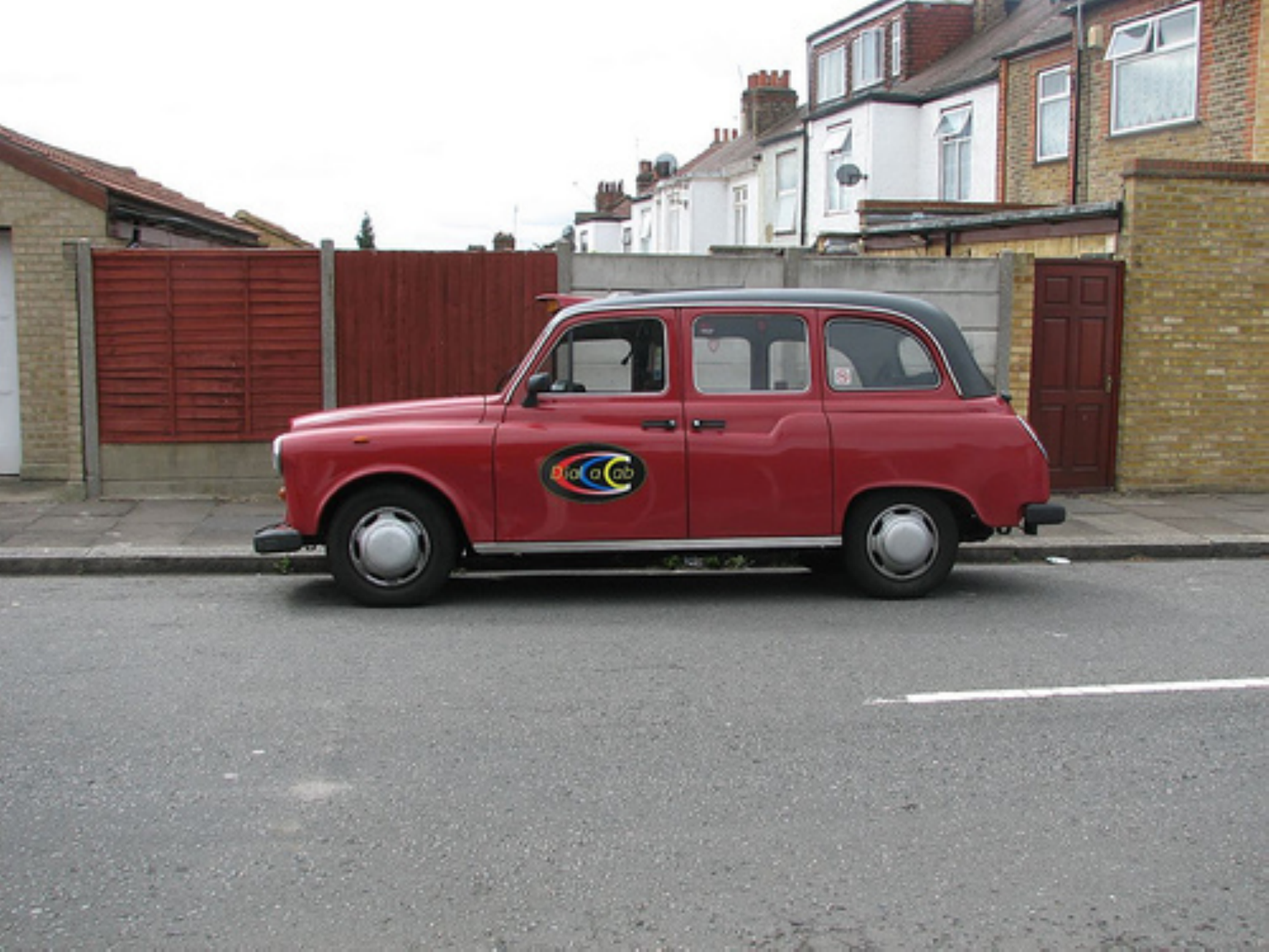}\caption{original image}\label{orig_3}
            \end{subfigure} &
            \begin{subfigure}{.16\textwidth}
                \includegraphics[width=\linewidth]{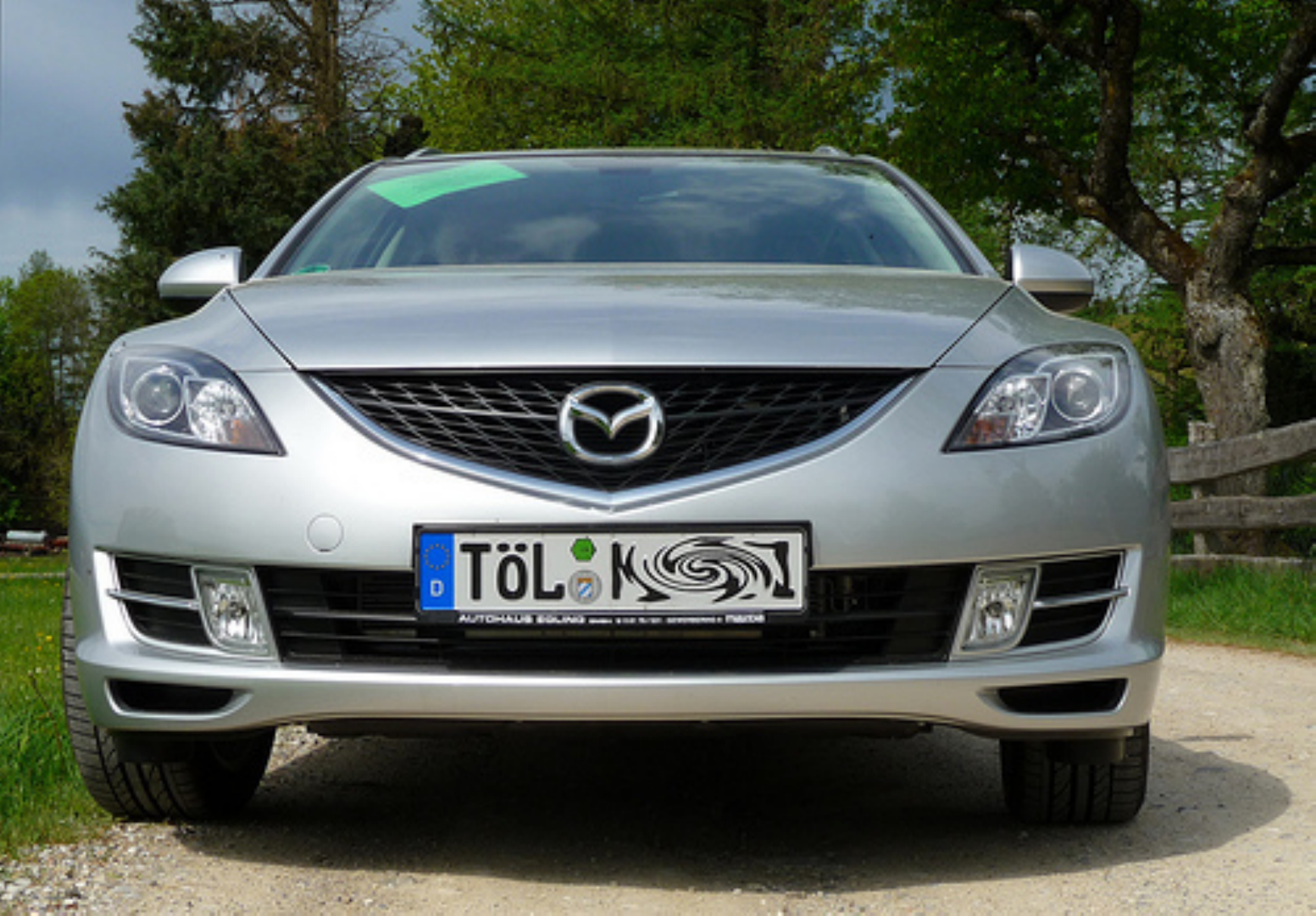}\caption{original image}\label{orig_4}
            \end{subfigure}
            \\
             \cmark \ \textit{car} & \xmark \ \textit{not car} & \xmark \ \textit{not car}\\
             \midrule
             \cellcolor{orange!8}
             \begin{subfigure}{.16\textwidth}
             \includegraphics[width=\linewidth]{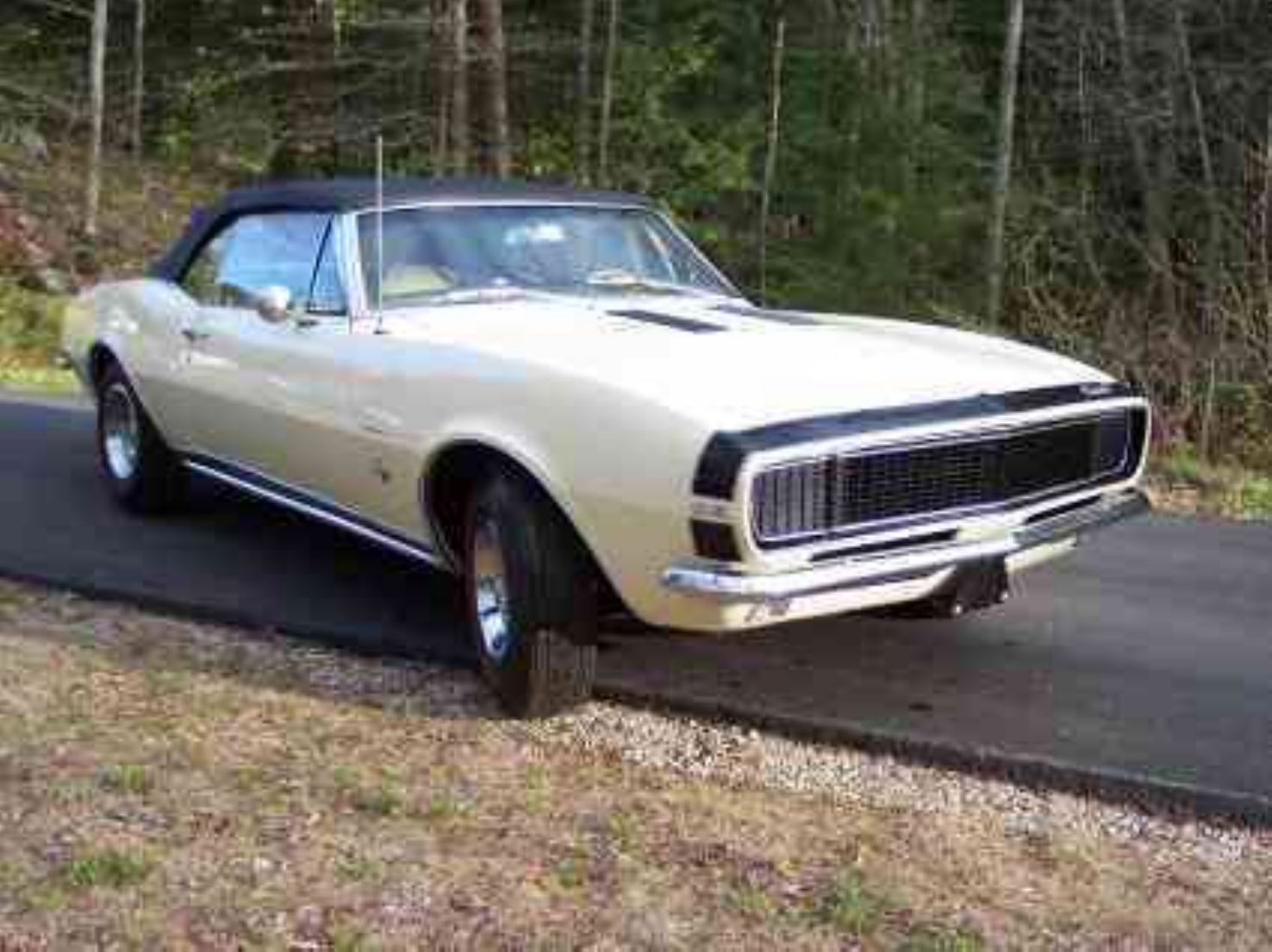}\caption{brightness}\label{brightness_small}
             \end{subfigure} &
             \cellcolor{orange!8}
             \begin{subfigure}{.16\textwidth}
                \includegraphics[width=\linewidth]{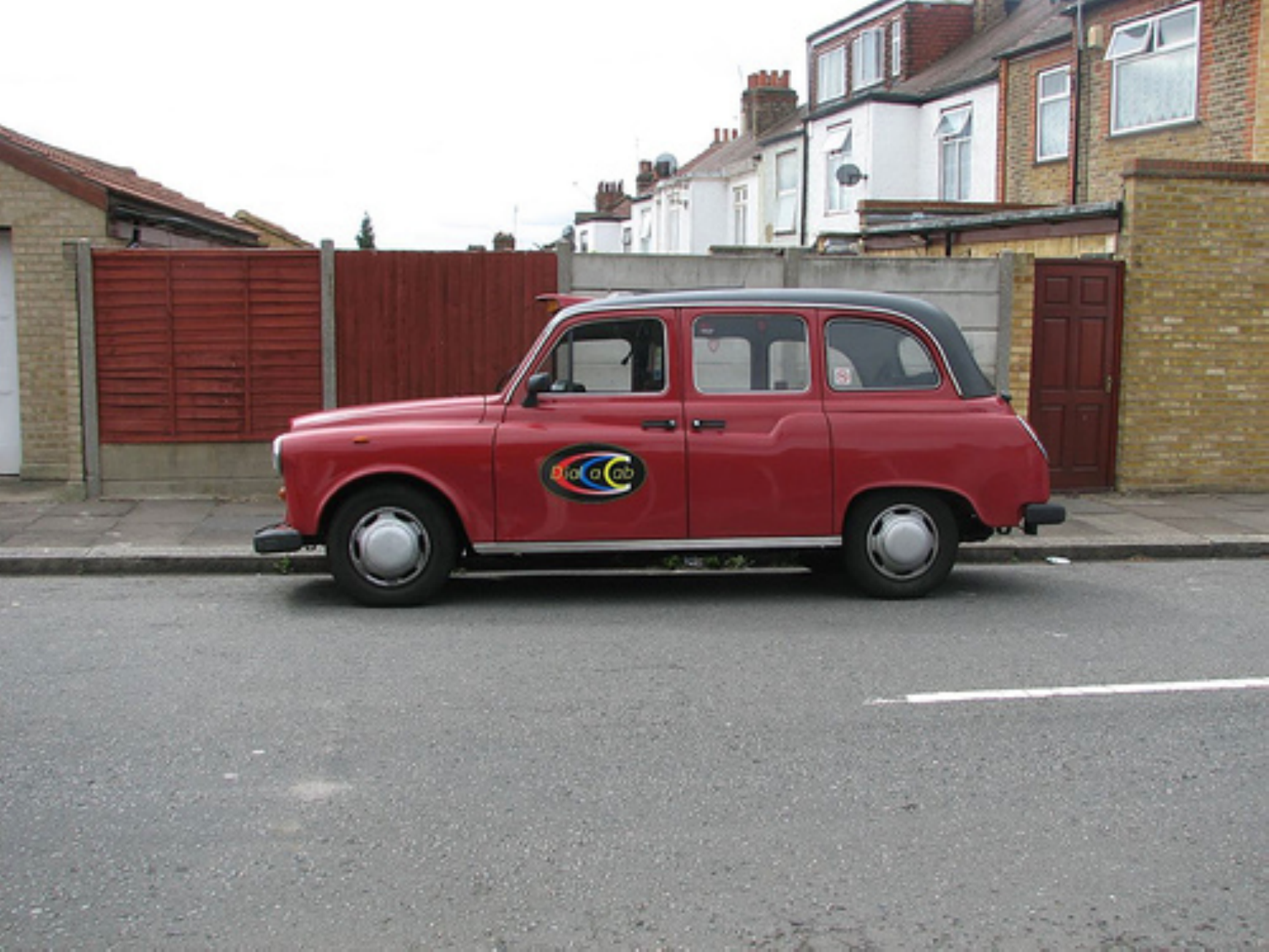}\caption{brightness}\label{brightness_small_1}
             \end{subfigure}  &
             \cellcolor{orange!8}
             \begin{subfigure}{.16\textwidth}
                \includegraphics[width=\linewidth]{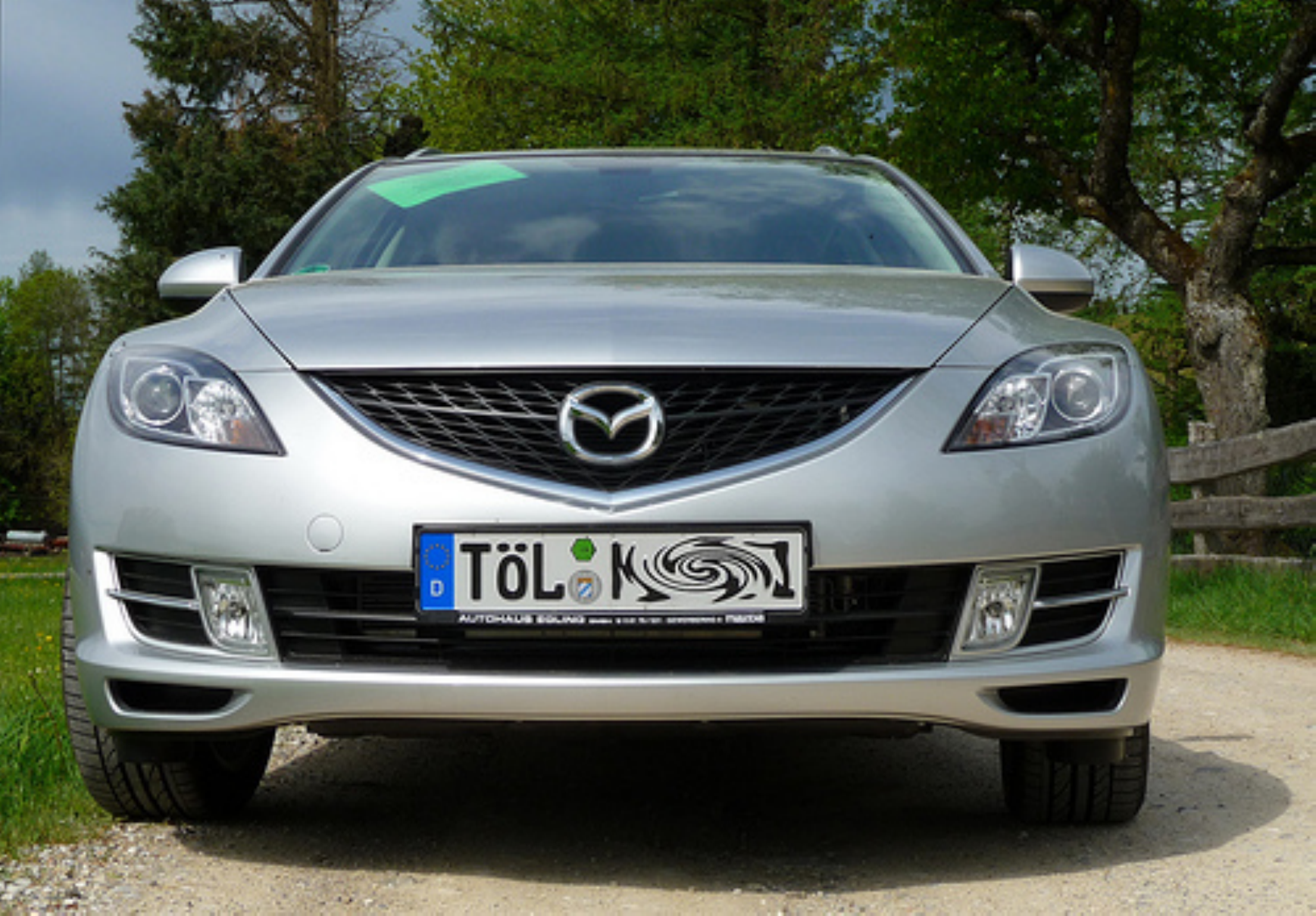}\caption{brightness}\label{brightness_small_2}
             \end{subfigure}\\
            \cellcolor{orange!8}\cmark \ \textit{car} &  \cellcolor{orange!8}\xmark \ \textit{not car} & \cellcolor{orange!8}\xmark \ \textit{not car}\\
            \midrule
             \cellcolor{orange!25}
             \begin{subfigure}{.16\textwidth}
             \includegraphics[width=\linewidth]{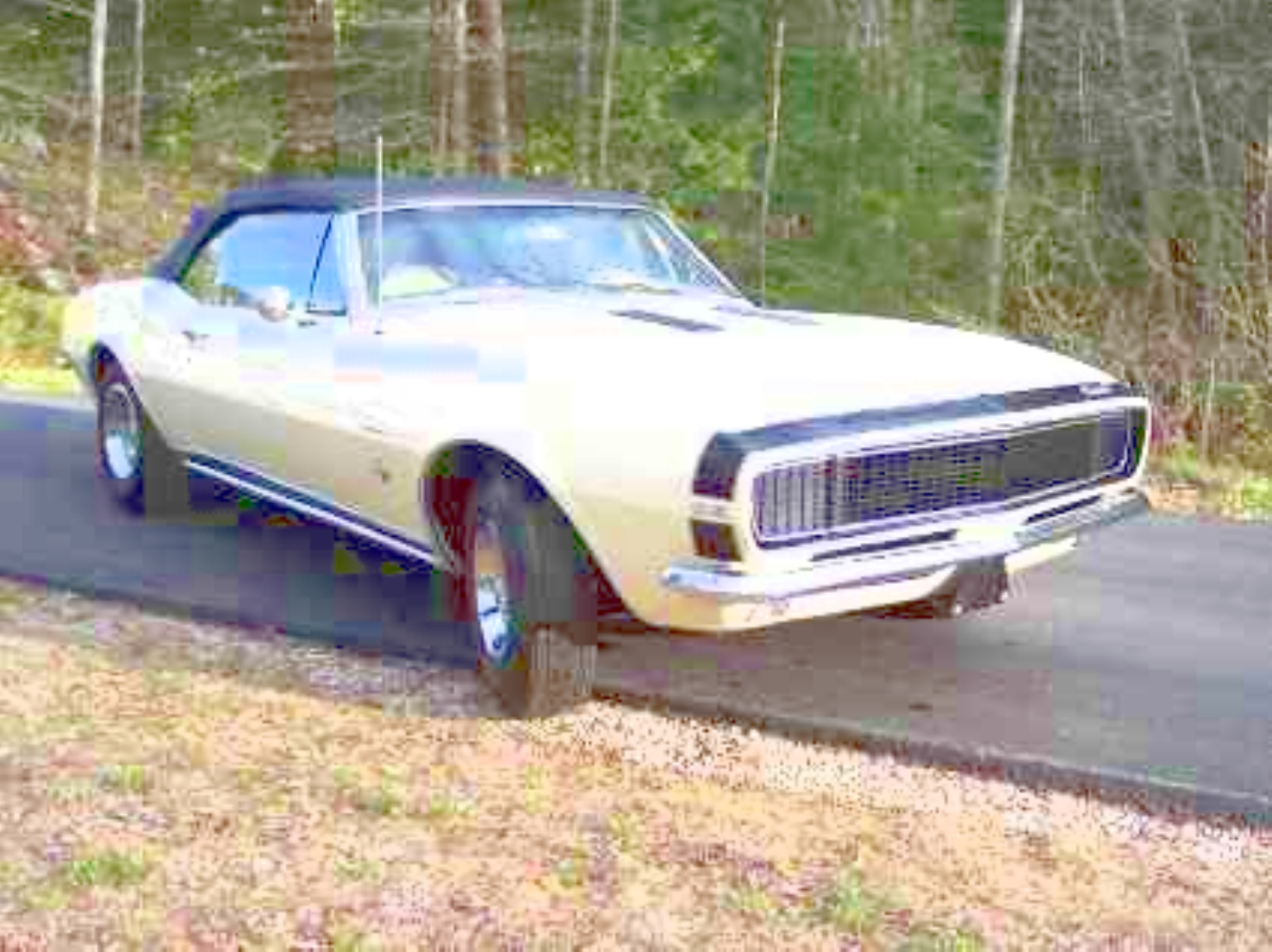}\caption{brightness ++}\label{brightness_big}
             \end{subfigure} &
             \cellcolor{orange!25}
             \begin{subfigure}{.16\textwidth}
                \includegraphics[width=\linewidth]{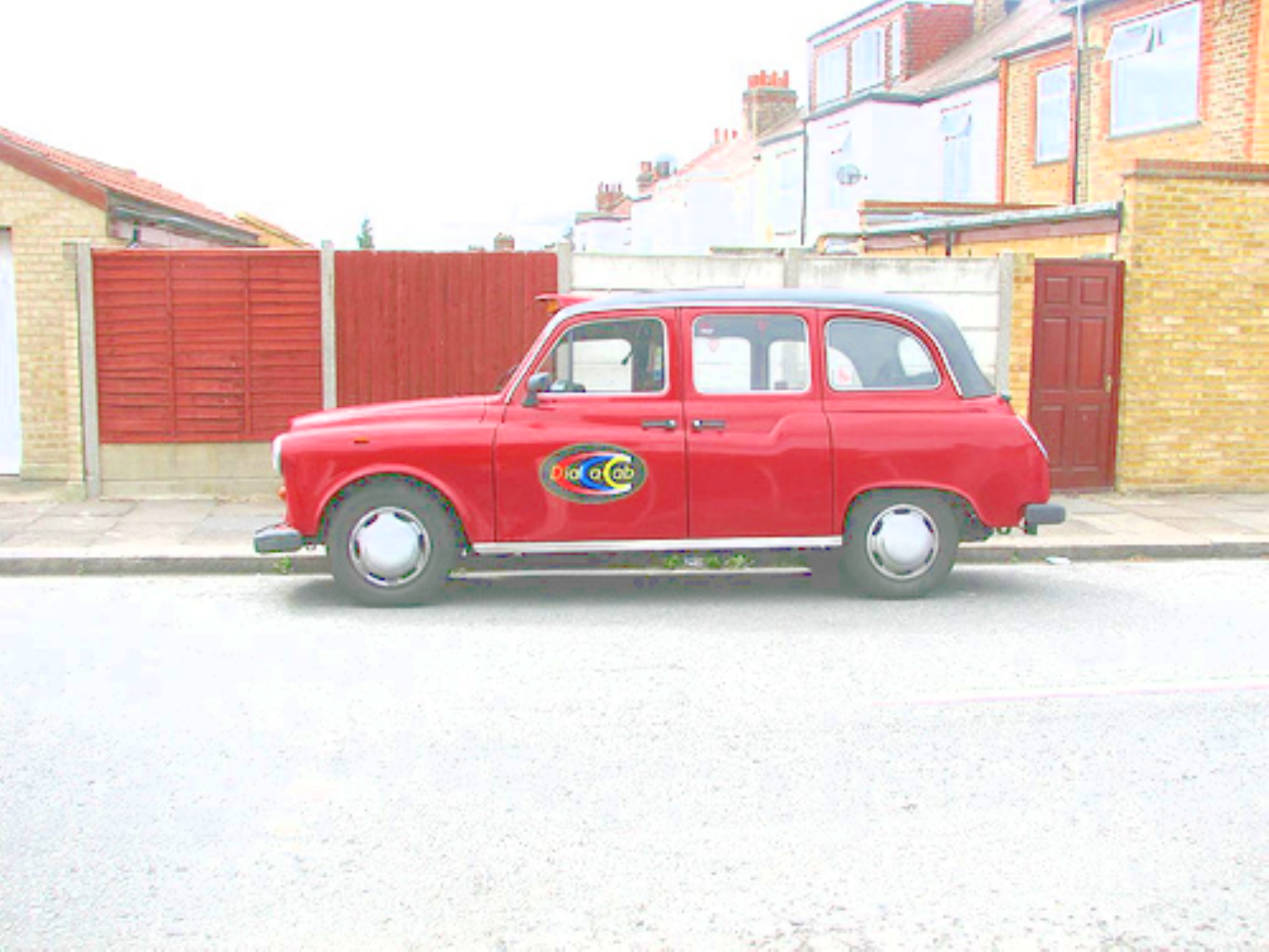}\caption{brightness ++}\label{brightness_big_1}
             \end{subfigure}  &
             \cellcolor{orange!25}
             \begin{subfigure}{.16\textwidth}
                \includegraphics[width=\linewidth]{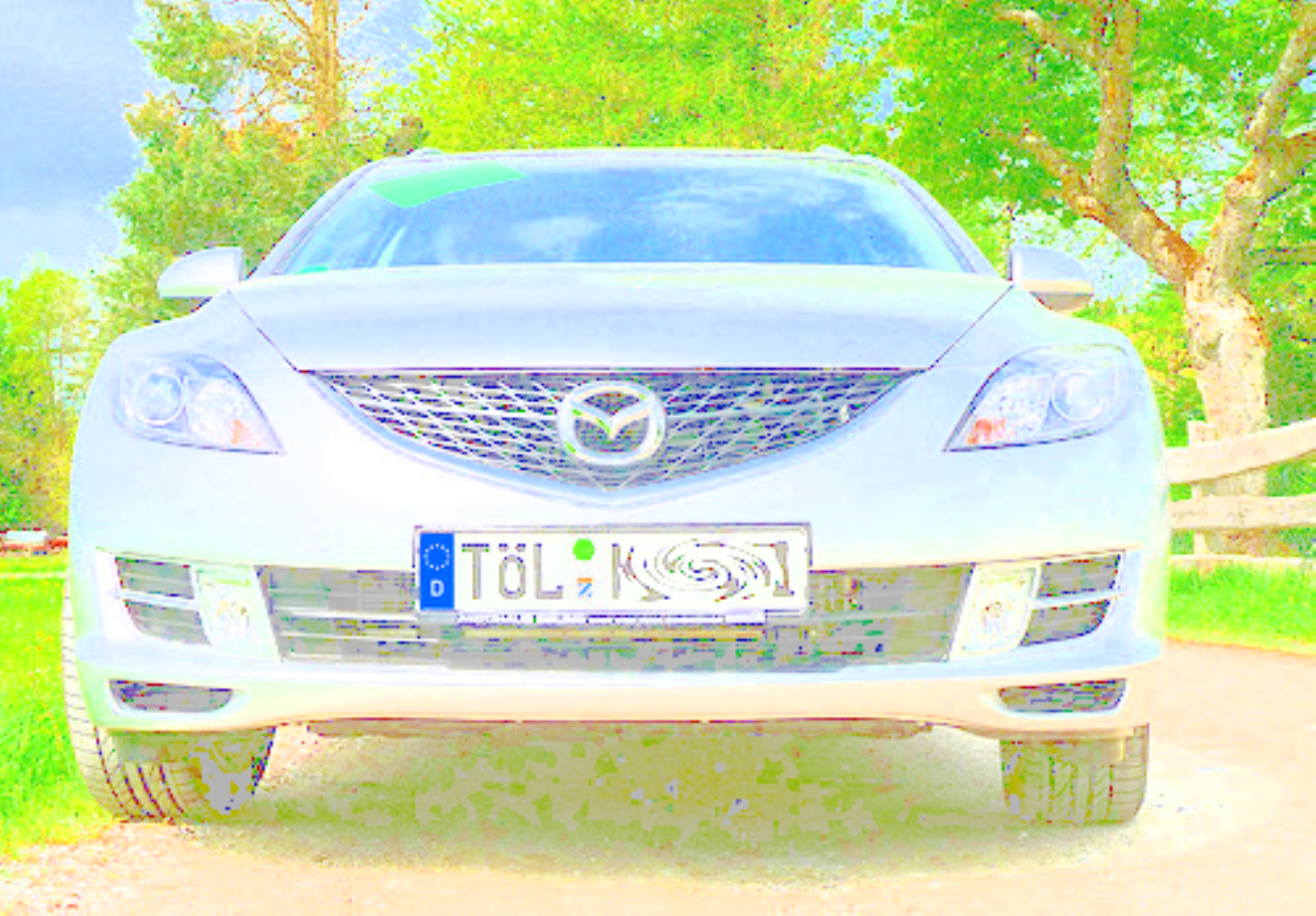}\caption{brightness ++}\label{brightness_big_2}
             \end{subfigure}\\
            \cellcolor{orange!25}\cmark \ \textit{car} &  \cellcolor{orange!25}\xmark \ \textit{not car} & \cellcolor{orange!25}\cmark \ \textit{car}\\ 
              \bottomrule
     \end{tabular}    
}
    \caption{{\small Image recognition on original and transformed images.
    Images from (d) to (i) display the same transformation, brightness, applied with different magnitudes. 
    The classification result of ResNet50 is shown in italics under each image.
    }}\label{fig:req-example}
    \vspace{-0.2in}
\end{figure}

\begin{mytheo}
{Correctness-preservation requirement,\\ with parameters $T_X$ and $t_{c}$}{}\label{def:cp}

\textbf{Intuitively}: For the range of changes in images that do not affect human performance ($\Delta_v \leq t_{c}$), the \emph{performance} of machine vision component $f$ should not be affected as well. Note that ground truth \emph{is} required.

\textbf{Formally}:
We require the performance $m$ of $f$ for transformed images to be equal to or larger than that for original images: $m( f , f^*, P_{T_{X,t_c}})\geq m(f,f^*,P_X)$.
\end{mytheo}
\noindent
\textbf{Example:} For the transformation brightness,
\emph{correctness-preservation} requires $m_{t_c}=m(f, f^*, P_{T_{X,t_c}})$, the classification accuracy of ResNet50 on all transformed images  (which is the percentage of \emph{correct} predictions in Fig.~\ref{brightness_small}-\ref{brightness_big_2}), to be at least $m_0= m(f,f^*,P_X)$, the classification accuracy of the model on all original images, which is the percentage of \emph{correct} predictions in Fig.~\ref{orig_2}-\ref{orig_4}. Both accuracies are $1/3 \approx 33\%$, and the requirement is satisfied.

We then introduce our \emph{prediction-preservation} requirement class. Given a distance measure $d(f(x), f(x'))$, which measures distance between the output of $f$ on two input images, we define a \emph{prediction-similarity} measure $s(f,P_{X\times X})=1-E_{(x, x')\sim P_{X\times X}}[d(f(x), f(x'))]$, which measures the expected similarity between the output of $f$ on two images drawn from $P_{X\times X}$, a distribution of image pairs. Note that $d$ should also be adequate for the vision task; for example, for image classification, $d(f(x), f(x')) = 0$ if $f(x) == f(x')$ and $1$ otherwise. We define the distribution of \emph{pairs} of original and transformed images that are within the human-tolerated range for prediction-preservation $\Delta_v \leq t_p$ by conditioning the joint distribution $P_{T_{X}}$ as follows:  
$P_{T_{X, t_p}}(x,x')= P_{T_X}(x, x' |  \Delta_v(x, x') \leq t_p)$. Since $s$ compares outputs with the original outputs, $s$ of original images would always be 1, which is not necessarily achievable. As an alternative, we estimate $s$ of original images with $s$ of images with minimal image changes ($\Delta_v\leq\epsilon$). More precisely, we rank the image pairs by $\Delta_v$, determine $\epsilon$ as a lower $q$-th quantile in the ranking, and define the distribution of image pairs with $\Delta_v\leq\epsilon$ as $P_{T_{X, \epsilon}}(x,x')= P_{T_X}(x, x' |  \Delta_v(x, x') \leq \epsilon)$.

\begin{mytheo}{Prediction-preservation requirement, \\with parameters $T_X$ and $t_{p}$}{}\label{def:pp}

\textbf{Intuitively}: For the range of changes in images that do not affect human predictions ($\Delta_v\leq t_{p}$), the \emph{predictions} of machine vision component $f$ should stay unaffected as well.
Note that ground truth \emph{is not} required.

\textbf{Formally}: 
We require the prediction similarity $s$ of $f$ for all transformed images to be equal to or larger than that of images transformed with $\Delta_v \leq \epsilon$, which is: $s(f, P_{T_{X,t_p}})\geq s(f,P_{X,\epsilon})$.

\end{mytheo}
\noindent
\textbf{Example:} 
For the transformation brightness,
\emph{prediction-preservation} requires $s_{t_p}= s(f, P_{T_{X,t_p}})$, the prediction similarity of ResNet50 for all transformed images vs. originals, which is the percentage of predictions in images in Fig.~\ref{brightness_small}-\ref{brightness_big_2} \emph{preserved} from images in Fig.~\ref{orig_2}-\ref{orig_4} and thus $5/6\approx 83\%$, to be
at least equal to $s_0=s(f,P_{X,\epsilon})$, the prediction similarity of the model for images transformed with $\Delta_v \leq \epsilon$. Given the very small sample, we set $\epsilon$ to the median, and thus $s_0$ is the percentage of predictions preserved for images in Fig.~\ref{brightness_small}-\ref{brightness_small_2}, and $s_0=3/3=100\%$. Thus, the requirement is not satisfied.

Definitions of the two reliability requirements are similar, with two main differences. First, correctness-preservation relies on a performance metric to compare predictions to ground truth, whereas prediction-preservation uses prediction similarity to compare predictions on transformed images vs. originals. Second, correctness-preservation compares performance on the full range of transformed images with that on the originals, whereas prediction-preservation compares the prediction similarity for the full range of transformed images vs. originals to that for the minimally transformed images (i.e., $\Delta_v\leq\epsilon$) vs. originals. The design choices for the prediction-preservation requirement completely remove the need for human labels on test images, and make this requirement applicable in environments where such labels are unavailable.

Finally, we define \emph{reliability distance} as the difference between the target and the actual correctness- or prediction-preservation, i.e., $\Delta m = m_0-m_{t_c}$ and $\Delta s = s_0-s_{t_p}$, respectively. This distance indicates how well the MVC satisfies the respective requirement: zero distance indicates just meeting it; negative distance indicates performing better than the target by a margin; and positive distance indicates how far the MVC is from meeting the target.

%% file: estimating.tex
\pdfoutput=1
\vspace{-0.05in}
\section{Instantiating reliability requirements}
\label{estimate}
To obtain the reliability requirements range parameters  $t_c$, and $t_p$ in Defs.~2 and 3, we perform two experiments with human participants and then estimate the parameters from the experimental results to obtain threshholds at which the human performance drops statistically significantly (step~\RN{1}.a of Fig.~\ref{fig:flowchart}). 
This section first presents the experimental setup and procedure, and then  introduces our method for instantiating the requirements from the experimental results (requirement instantiation, steps~\RN{1}.a-b). 

\vskip 0.05in
\noindent
{\bf Experiments with human participants.}
The objective of the human experiments, one per dataset, is to obtain human predictions on original and transformed images, to be used to estimate $t_c$, and $t_p$ in step~\RN{1}.a.
The experiment inputs are the task to be preformed; the transformation $T_X$; the dataset of original images $\{x_i\}$, $x_i\sim P_X$, with their ground-truth labels $\{f^*(x_i)\}$; and distributions $P_C$ for each transformation parameter. Given these inputs, we generate a sample of original-transformed images $\{(x_i,x_{i,j}’)\}$, $(x_i,x_{i,j}’)\sim P_{T_X}$, by randomly selecting $x_i$ from $\{x_i\}$ and $c_j\sim P_C$, and transforming $x_{i,j}’=T_X(x_i,c_j)$.
To obtain the human predictions for each image in $\{(x_i,x_{i,j}’)\}$ for image classification, we follow the experimental design of Geirhos et al.~\cite{Geirhos2018GeneralisationIH}. The experiment is a \emph{forced-choice image categorisation task}: humans are presented with the images with transformations applied, for 200 ms, and asked to choose one of the two categories (e.g., car or not car). Between images, a noise mask is shown to minimize feedback influence in the brain~\cite{Geirhos2018GeneralisationIH}.
The tasks are timed to ensure fairness in the comparison between humans and machine~\cite{Firestone26562}. However, in contrast to the work by Geirhos et al., we ensure that the full range of achievable $\Delta_v$ values is covered when sampling from $P_C$, and we also collect human predictions for the original images, so that we can estimate prediction preservation. A given subject is never shown more than one version of $x_i$, whether original or transformed. Note that human predictions for originals are different from their ground truth labels: labelers take as long as needed per image to classify it, but our subjects have only 200 ms to see each image.
The human data is specific to a task, an image distribution, and a class of transformations, and thus has to be collected for the combination of the three. 
    In other words, the data is reusable for different samples from the same distribution, or, intuitively, for images sharing the same characteristics, e.g., the same image resolution, the same objects, etc. For example, it is reusable across different sets of images of road scenes taken within the same geographic area using cameras with same the resolution and image quality.

To generate predictions for our evaluation, we perform the experiment on two datasets: 
ILSVRC'12~\cite{ILSVRC2012} and CIFAR-10~\cite{krizhevsky-09}. While CIFAR-10 has images of much lower resolution than ILSVRC'12, we include this dataset to compare our method to the existing work on robustness checking, which uses CIFAR-10.
We also select the eight safety-related transformations (see images from Fig.~\ref{fig:motivating}c-j), as discussed in Sec.~\ref{transformations}, and set $P_C$ to be uniform. To fit our labeling budget, we limit the task to a binary classification problem of recognizing car instances. Also, while we apply the eight transformations to ILSVRC'12, we limit the experiment on CIFAR-10 to four transformations that are also used in the works we compare with.
To differentiate between car and non-car instances, we use the class hierarchy from the ILSVRC'12 dataset.
For each of the selected transformations, we sample 1000 pairs $(x_i,x_{i,j}’)$, and have each image (original or transformed) labeled by five humans.
To achieve this, we divide the 1,000 (pair samples) $\times$ 8 ($\#$considered) transformations into batches of 20 images.
Each batch is shown five times to different participants using the platform Amazon Mechanical Turk.
We include qualification tests and sanity checks aimed to filter out participants  misunderstanding the task and spammers~\cite{Papadopoulos-et-al-2017}, and
only consider results from participants who pass both tests.
As a result, for the ILSVRC'12 image classification task, we use $\{x_i\}$ with 13,000 car images and same number of non-car images, and collect human predictions for 40,000 ($= 5 \times 8 \times 1,000$) transformed images and the same number of the original images, for a total of 80,000 predictions. Note that the effort required for measuring human performance is significantly smaller than the the dataset labeling effort needed for model training. For example, we collect human predictions for 5,000 transformed images per transformation, with 0.2\,s timebox per image, for a total of 1000\,s.  Training sets are typically over 100,000 images, with at least three ground truth labels assigned independently to each image (for quality control) and each takes multiple seconds; e.g., 100,000 x 3 x 2\,s = 600,000\,s.
The human experiment results can be found in supplementary material$^{\ref{gitlink_note}}$.

\vskip 0.05in
\noindent
{\bf Estimating tolerated range parameters and instantiating requirements.}
We propose the following procedure to estimate $t_c$ and $t_p$ from the experimentally-obtained human predictions (step~\RN{1}.a).
The key idea is to group and order the image pairs by $\Delta_v$, compute the human performance $m_k$ and human prediction similarity $s_k$ in each group, and use a statistical test to determine the $t_c$ [resp. $t_p$] value of $\Delta_v$ at which $m_k$ [resp. $s_k$] drops significantly from the human performance $m_0$ for the original images  [resp. the human prediction similarity $s_0$]. Recall that $s_0$ is the the human prediction similarity for images transformed with $\Delta_v\leq \epsilon$ vs. originals; we set $\epsilon$ to the lower 5th percentile.

More precisely, we determine threshold $t_c^{(l)}$ and $t_p^{(l)}$ for a given transformation $T_{X,l}$ from the image pairs $(x_i,x_{i,j}’)$, the human predictions for these images, and their ground truth $f^*$. First, we compute $\Delta_v$ for each pair, and sort the pairs by their $\Delta_v$ into $r$ intervals, defined by $r+1$ equidistanced thresholds $t_k$, with $t_0=0$ and $t_r=1$. We then process the result using smoothing splines~\cite{koenker-94} to reduce randomness and remove outliers. 
Then, to estimate $t_c^{(l)}$, for each $k\in [1..r]$ we compute the probability $p_k$ that $m_k$ on the transformed images in the $k$-th interval $[t_{k-1},t_k]$ is below $m_0$. This probability is obtained using the single-sided binomial test. We then determine the interval with the smallest $t_k$ for which $p_k \geq 0.05$, and return this $t_k$ as $t_c^{(l)}$. Similarly, for $t_p^{(l)}$, we compute the probability $p_k$ that $s_k$ for the original-transformed image pairs in $[t_{k-1},t_k]$ is below $s_0$. Then $t_c^{(l)}$ is the smallest $t_k$ for which $p_k \geq 0.05$.

With the above procedure, we can now estimate the parameters for the task of recognizing cars  for each of the eight selected transformations (Fig.~\ref{fig:motivating}) using our experimental results.  The instantiated parameters are shown in Tbl.~\ref{tab:req_inst}. 
 
With these parameters, we obtain the instantiated machine-verifiable reliability requirements for each transformation (step~\RN{1}.b). For example, for the transformation brightness, given an MVC that recognizes cars in images, the \emph{correctness-preservation} requirement says that the MVC's recognition accuracy should not decrease if the visual change in the images is within the range $\Delta_v \leq 0.8$; and the \emph{prediction-preservation} requirement says that the percentage of labels humans can preserve after a brightness change should not decrease if the visual change in the images is within the range $\Delta_v \leq 0.86$. To obtain the instantiated requirements for other transformations, only the parameter values need to be replaced with the estimated values in Tbl.~\ref{tab:req_inst}.

\begin{table}[t]
\small
\centering
  \caption{{\small Estimated parameters for correctness- ($t_{c}$) and prediction-preservation ($t_{p}$) requirements using human experiment results for the task of recognizing car instances.}}
  
  \label{tab:req_inst}
  \scalebox{1}{
  \begin{tabular}{|c|c|c |c||c|c|c|}
    \hline
  \cellcolor{gray!25}& \cellcolor{gray!25}\shortstack{transformation} & \cellcolor{gray!25}\shortstack{$t_{c}$} & \cellcolor{gray!25}\shortstack{$t_{p}$}& \cellcolor{gray!25}\shortstack{transformation} & \cellcolor{gray!25}\shortstack{$t_{c}$} & \cellcolor{gray!25}\shortstack{$t_{p}$}\\

    \hline
      \parbox[t]{2mm}{\multirow{4}{*}{\rotatebox[origin=c]{90}{Imagenet}}} & RGB&  0.82&  0.67 & brightness & 0.87 & 0.87\\
    \cline{2-7}
    &contrast & 0.77 &  0.28 & Gaussian noise & 0.91& 0.91\\
    \cline{2-7}
    &defocus blur&  0.98 & 0.94 & color jitter & 0.48 & 0.48\\
    \cline{2-7}
    &frost&  0.84 &  0.91 & \shortstack{jpeg compression} & 0.94 & 0.94\\
    \hline
    \hline
    \parbox[t]{4mm}{\multirow{2}{*}{\rotatebox[origin=c]{90}{\small \shortstack{CIFAR\\-10}}}}&brightness&  0.78 & 0.89 & contrast & 0.63 & 0.86\\
    \cline{2-7}
    &frost&  0.61 &  0.61 & \shortstack{jpeg compression} & 0.60 & 0.60\\

    \hline
  \end{tabular}
  }
  \vspace{-0.1in}
\end{table}

%% file: testing.tex
\pdfoutput=1
\vspace{-0.1in}
\section{Checking reliability requirements}
\label{testing}
In this section, we describe a method for automatically checking whether MVCs satisfy our machine-verifiable requirements (see steps~\RN{2}.a-c \emph{requirement checking} in Fig.~\ref{fig:flowchart}). 
Requirement checking takes as inputs a list of images, a set of transformations, and an MVC under validation.  
It generates test cases (step~\RN{2}.a) within the specified range of $\Delta_v\leq t_c$ or $t_p$; runs the tests on the MVC (step~\RN{2}.b); and checks whether the MVC satisfies the requirements by estimating the reliability distance (step~\RN{2}.c).

We test the requirements satisfaction by estimating MVC performance or prediction preservation through sampling. This is necessarily, since we do not have direct access to $P_X$ but only its samples.  
Our test case generation is based on the  \emph{bootstrap method}~\cite{efron1994introduction}, which estimates the metrics $m_0$, $m_{t_c}$, $s_0$, and $s_{t_p}$ through sampling the test data with replacement. Since these metrics are defined as expected values or means, by Central Limit Theorem, the values of these metrics computed for sample batches, denoted for each batch $i$ as $\bar{m}_{0,i}$, $\bar{m}_{t_c,i}$, $\bar{s}_{0,i}$, and $\bar{s}_{t_p,i}$, respectively, are normally distributed. Following the bootstrap method, we obtain the population estimates by computing
the means $\hat{m}_0$, $\hat{m}_{t_c}$, $\hat{s}_0$, and $\hat{s}_{t_p}$
and the standard deviations
$\sigma_{\hat{m}_0}$, $\sigma_{\hat{m}_{t_c}}$, $\sigma_{\hat{s}_0}$, and $\sigma_{\hat{s}_{t_p}}$
of the respective batch value sets
$\{\bar{m}_{0,i}\}$, $\{\bar{m}_{t_c,i}\}$, $\{\bar{s}_{0,i}\}$, and $\{\bar{s}_{t_p,i}\}$.
To do so, for each transformation $T_X$ and a list of original images $X$, we sample $n$ batches of $k$ images from $X$ and then generate a transformed image by applying $T_X$ to each sampled original image with randomly sampled parameter values while ensuring the required $\Delta_v$ range, i.e., $n \times k$ pairs in total. Since sampling is part of the process, $n$ and $k$ should be determined based on $|X|$, and the bigger they are, the more accurate the estimated results would be~\cite{efron1994introduction}. Although the lower-bound numbers for $n$ and $k$ are hard to determine, one can check whether the sampling is sufficient as the bootstrap method always converges with enough batches of samples for normal distributions~\cite{bootstrap}. A choice of $n$ is considered sufficiently large if two separate runs with different random seeds result in similar estimated values. 
After generating the tests, our method runs them on the MVC under validation, and obtains the MVC predictions for all the original images and for each batch $i$ of transformed images. We then compute the sample batch estimates of the four metrics, i.e., $\{\bar{m}_{0,i}\}$, $\{\bar{m}_{t_c,i}\}$, $\{\bar{s}_{0,i}\}$, and $\{\bar{s}_{t_p,i}\}$, and take mean ($\Delta \hat{m},\Delta \hat{s}$) and standard deviation ($\sigma_{\Delta \hat{m}}, \sigma_{\Delta \hat{s}}$) of each set as the population estimates.

Finally, we want to show that the reliability distance for each requirement is zero or negative, i.e., $\Delta m\leq 0$ for correctness-preservation and $\Delta s\leq 0$ for prediction-preservation.
Since our estimates from the previous step
are normally distributed,
their differences are also normally distributed. Thus, the reliability
distance estimates have
the following means and
standard deviations:
$\Delta \hat{m}=\hat{m}_0-\hat{m}_{t_c}$ and $\sigma_{\Delta \hat{m}}=\sqrt{\sigma_{\hat{m}_0}^2+\sigma_{\hat{m}_{t_c}}^2}$; and
$\Delta \hat{s}=\hat{s}_0-\hat{s}_{t_p}$ and $\sigma_{\Delta \hat{s}}=\sqrt{\sigma_{\hat{s}_0}^2+\sigma_{\hat{s}_{t_p}}^2}$.
To ensure that the reliability distances
are zero or negative with a confidence
$1-\alpha=95\%$, we use the right-handed
confidence interval.
Thus, $\Delta m\leq 0$ with confidence $1-\alpha$ iff 
$\Delta \hat{m} + z_{\alpha}\sigma_{\Delta \hat{m}}\leq 0$, where $z_{\alpha}$ is the z-value corresponding to an area $\alpha$ in the right tail of a standard normal distribution, with $z_{0.05}=1.645$ for $95\%$ confidence.
Similarly,
$\Delta s\leq 0$ with confidence $1-\alpha$ iff 
$\Delta \hat{s} + z_{\alpha}\sigma_{\Delta \hat{s}}\leq 0$.

For example, to check whether ResNet50 satisfies our instantiated requirements for the transformation Gaussian noise (see Tbl.~\ref{tab:req_inst}) for the task of recognizing cars,
the testing method first generates tests with the original and the transformed images within the $\Delta_v$ range specified in the requirements.
The original images are sampled from the ILSVRC'12 validation dataset using bootstrap with $n=200, k=50$ and the Gaussian noise transformation. 
Then we run the generated tests on ResNet50; compute the four sets of metrics $\{\bar{m}_{0,i}\}$, $\{\bar{m}_{t_c,i}\}$, $\{\bar{s}_{0,i}\}$, and $\{\bar{s}_{t_p,i}\}$ over the batches; and then compute $\Delta \hat{m}=0.0045$, $\sigma_{\Delta \hat{m}}=0.0061$,
$\Delta \hat{s}=0.0011$, and $\sigma_{\Delta \hat{s}}=0.0045$.
We check for correctness-preservation with $95\%$ confidence: $\Delta \hat{m} + z_{0.05}\sigma_{\Delta \hat{m}}>0$; and prediction-preservation with $95\%$ confidence: $\Delta \hat{s} + z_{0.05}\sigma_{\Delta \hat{s}}>0$.
Therefore ResNet50 does not satisfy either of the requirements.
Note that by estimating the reliability distance, we provide engineers with a
quantitative measure of how much
improvement is needed to meet the requirements in case they are not met.

%% file: evaluation_new.tex
\pdfoutput=1
\section{Evaluation}
\label{sec:evaluation}
While our approach is defined for any computer-vision task, in this paper we demonstrate its feasibility on a particular domain: image classification, using parameters instantiated via human performance data collected for this domain as explained in Sec.~\ref{testing}.

First, we evaluate the generality of our instantiated image classification requirements.
For a specific transformation, our instantiated requirements contain the tolerated range of changes that do not affect human performance (see Sec.~\ref{requirements}), estimated from experiments with human participants.
Since such experiments are costly, we aim to minimize the number of experiments that need to be conducted.
To achieve this goal, we would like to reuse the collected human performance results for new sets of images from the dataset, different from the ones presented to the humans during the experiments. We expect the images to come from the same dataset to share the underlying data-generating distribution $P_X$.
Crucially, to be reusable, our requirement parameters should not be affected by the choice of the images included in the experiments with human participants.
Since our requirements are defined on a particular distribution of images, we aim to answer \textbf{(RQ1)}: How reusable are the thresholds $t_c$ and $t_p$ over different samples from the same image distribution?

Second, existing methods for evaluating reliability of image classification MVCs consider either small, imperceptible image changes or an arbitrary range of perceptible changes in images.
In this work, our focus is on a meaningful range of changes in images, the one that does not affect human vision, which includes both imperceptible and perceptible changes.
Since our goal is to use human performance as a baseline (i.e.,``if humans can see it, so should an MVC''), we are interested in understanding how well the existing reliability evaluation approaches already cover the human-tolerated range.
We are also interested in comparing the distribution of test cases generated by our method
(step~\RN{2}.a in Fig.~\ref{fig:flowchart}) with those from the other reliability methods, to see whether our method addresses the range better.
Therefore, we aim to answer
\textbf{(RQ2)}: How well do the existing reliability evaluation methods cover the human-tolerated range of changes?

Finally, we would like to determine whether checking the reliability of image classification MVC models with our method in the human-tolerated range of changes reveals reliability gaps in state-of-the-art image classification models.
To do so, we aim to answer \textbf{(RQ3)}:
How effective is our requirement checking method in identifying reliability gaps compared to existing approaches?

\vskip 0.05in
\noindent
{\bf RQ1. }
To answer RQ1, we compare human-tolerated ranges of transformations (parameters $t_{c}$ and $t_{p}$) estimated using our requirement instantiation method with \emph{different} sets of images from the ILSVRC'12 experiment.
We randomly selected two subsets of our results containing $60\%$ of all the images included in the experiment. We compared the similarity of the spline models obtained using these two subsets with all experiment results. As suggested by Koenker et al.~\cite{koenker-94}, two spline models are considered similar if their $83\%$ confidence intervals overlap. Following this, for each of the eight transformation included in our experiment, we compared the confidence intervals of the estimated spline models representing the two subsets and the entire set of experiment results. As a result, for all transformations, we observed that the spline models are unaffected. For example, the spline models obtained for the frost transformation are shown in Fig.~\ref{tab:splines}. Due to page limit, we include the plots and data in supplementary material$^{\ref{gitlink_note}}$. Since the parameters are derived using the spline models, unaffected spline models suggest that the parameters estimated are also unaffected. To conclude, different subsets of experiment results do not affect the parameters estimated. Therefore, we show evidence that our estimated human-tolerated ranges can be reused for images that are not included in the experiment with human participants, answering RQ1. Note that our requirements are defined on one image distribution, thus the thresholds cannot be reused for different image distributions. We can check this by comparing the values of $t_c$ and $t_p$ estimated using images from CIFAR-10~\cite{krizhevsky-09} and ILSVRC'12~\cite{ILSVRC2012}, shown in Tbl.~\ref{tab:req_inst} (Sec.~\ref{estimate}).

\begin{figure}
    \centering
    \includegraphics[width=\linewidth]{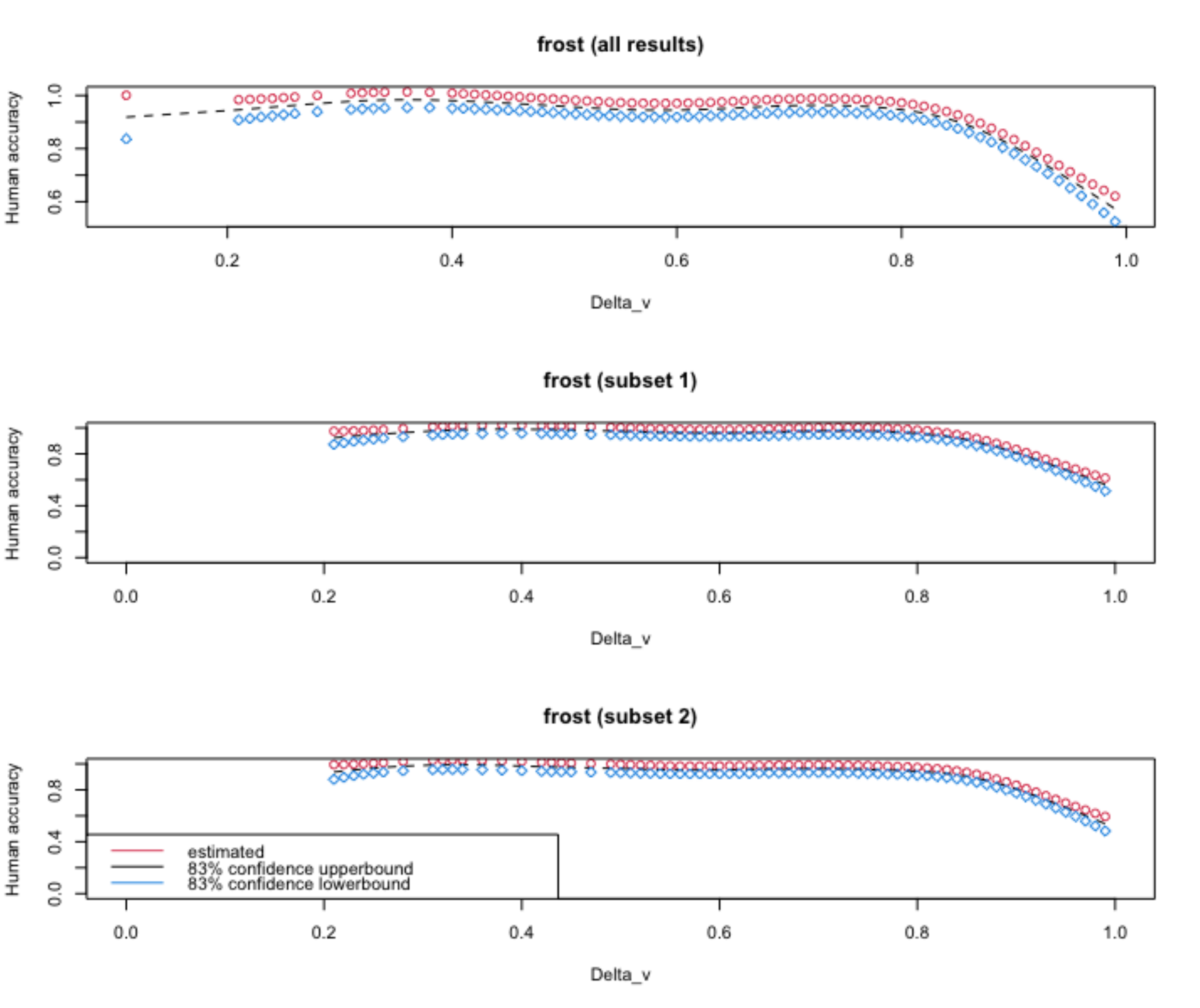}
    \vspace{-0.1in}
    \caption{A comparison of different subsets of experimental results for estimating $t_c$ for the frost transformation.}
    \label{tab:splines}
    \vspace{-0.2in}
\end{figure}

\vskip 0.05in
\noindent
{\bf RQ2.}
Existing methods for evaluating the reliability of MVCs with image transformations (imperceptible and perceptible changes) include \emph{metamorphic testing}~\cite{metamorphic_bugs, xie2011metatesting, Mekala_2019, zhang2018deeproad} and \emph{benchmarking}~\cite{hendrycks2019robustness}. Metamorphic testing is based on metamorphic relations; thus, instead of finding a range that does not affect human judgment, it consider all possible parameter values for each transformation included~\cite{metamorphic_bugs}. 
For this RQ, we compare with existing work that considers a broader range of changes: the state-of-the-art image corruption benchmark datasets Imagenet-c and CIFAR-10-c~\cite{hendrycks2019robustness}. 
These benchmark datasets include images transformed with five pre-selected parameter values for 19 arbitrarily chosen transformations.
Due to the low resolution of CIFAR-10 images, they look blurry to humans and thus do not share the same characteristics  with the ILSVRC'12 dataset~\cite{ILSVRC2012}. Therefore, to evaluate our method, we conduct an additional experiment with human participant using CIFAR-10~\cite{krizhevsky-09} images for four transformations (contrast, brightness, frost, and JPEG compression) and estimated the corresponding human-tolerated ranges, as described in Section~\ref{estimate}.
We answer RQ2 and RQ3 using six transformations considered by the other works:
brightness, contrast, defocus blur, frost, Gaussian noise, and jpeg compression.

\begin{figure}
\hspace{-0.1in}
\scalebox{0.95}{
    \begin{tabular}{m{0.24\textwidth} m{0.24\textwidth}}
    \begin{subfigure}{.24\textwidth}
      \centering
      \includegraphics[width=\linewidth]{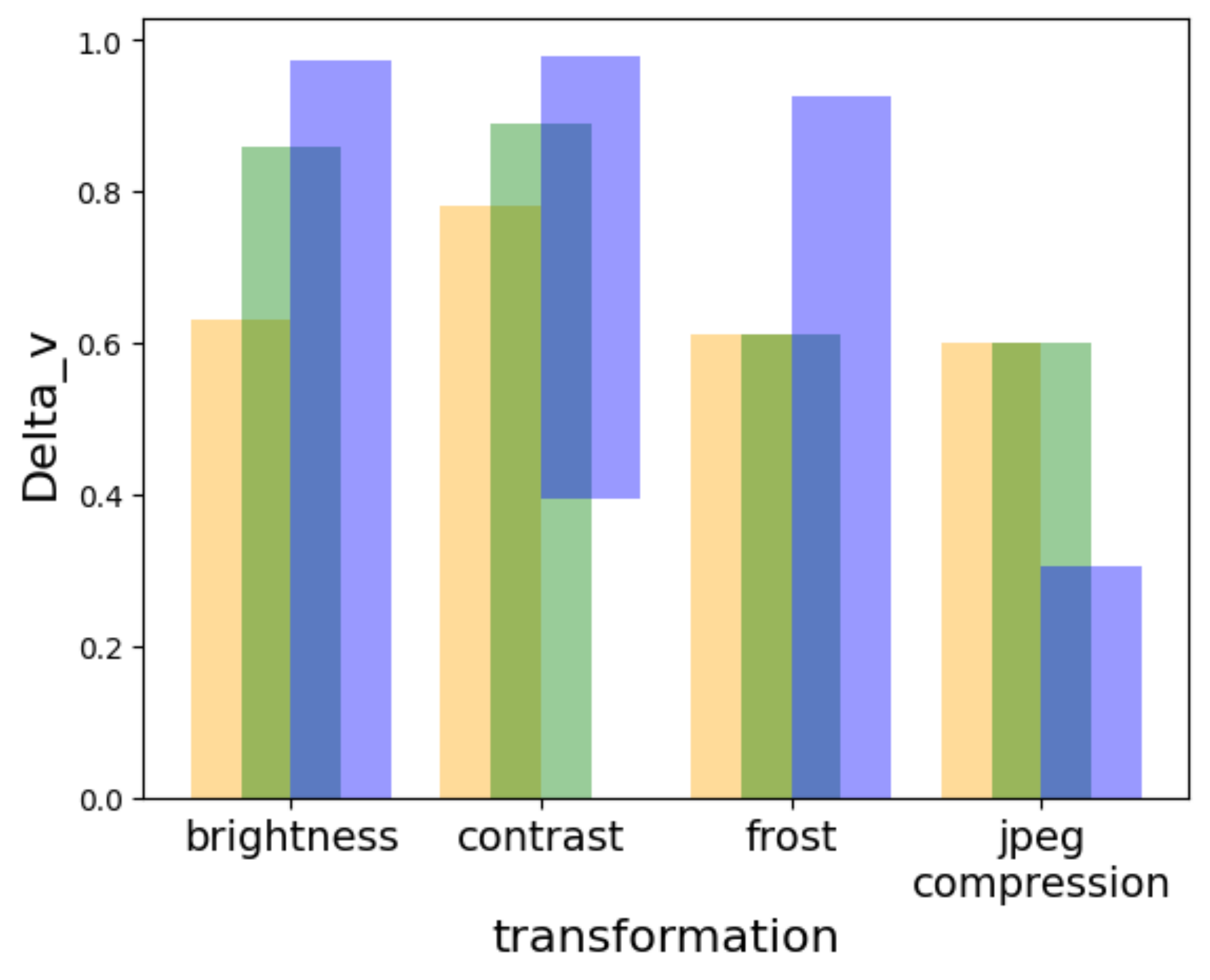}
      \caption{\small CIFAR-10-c and our ranges.}
      \label{fig:sub1} 
    \end{subfigure} &
    \begin{subfigure}{.24\textwidth}
      \centering
      \includegraphics[width=\linewidth]{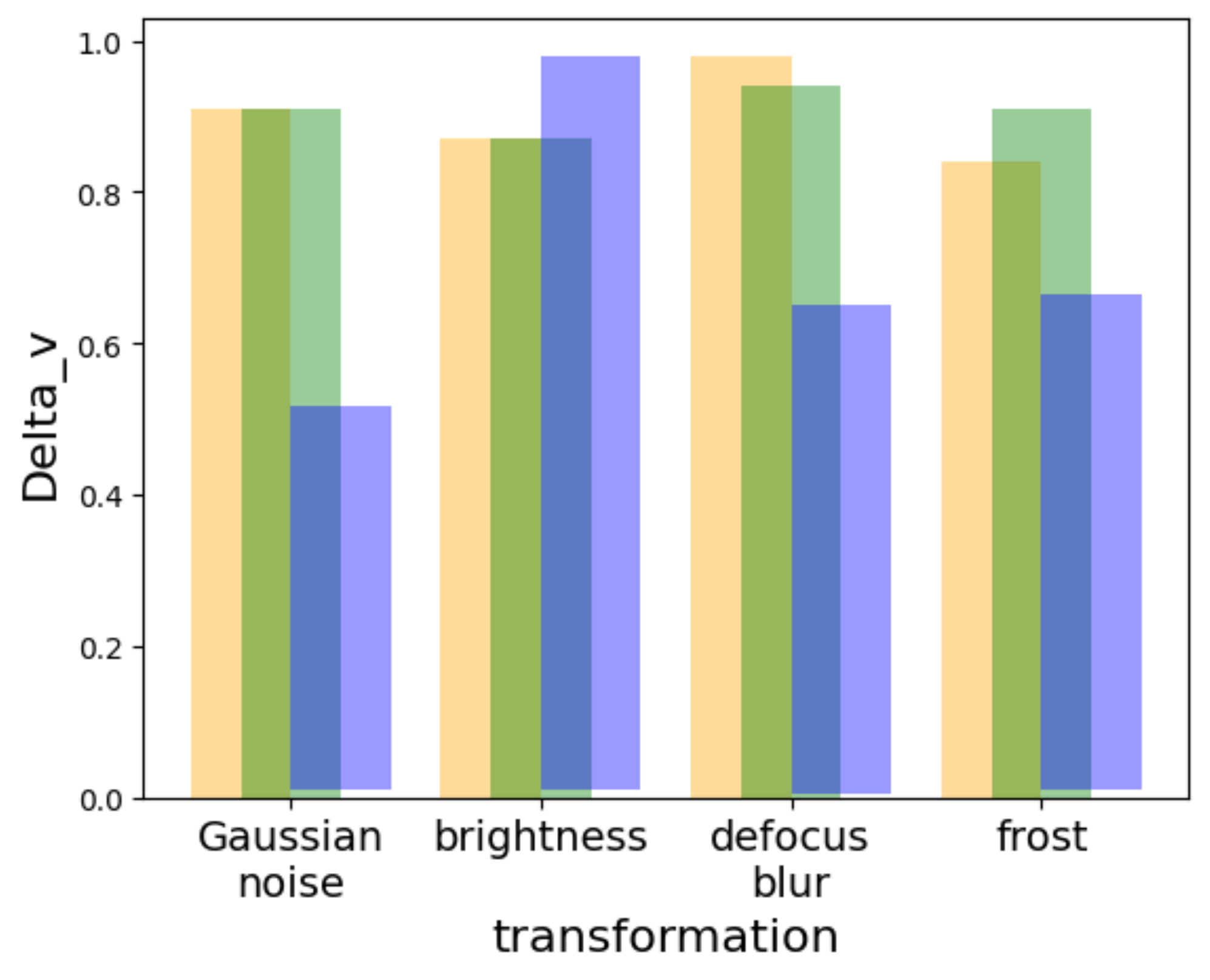}
      \caption{\small Imagenet-c and our ranges.}
      \label{fig:sub2}
    \end{subfigure}\\
    \hline
    \multicolumn{2}{|c|}{\shortstack{\textbf{Legend:} \includegraphics[scale=0.5]{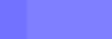} range of changes in CIFAR-10-c/imagenet-c,\\ \includegraphics[scale=0.5]{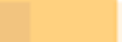} our range $[0, t_c]$, \includegraphics[scale=0.5]{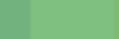} our range $[0, t_p]$}}\\
    \hline
    \end{tabular}
    }
    \caption{\small A comparison of our human-tolerated ranges for correctness-preservation and prediction-preservation requirements and the range of changes included in robustness benchmark datasets (Imagenet-c and CIFAR-10-c~\cite{hendrycks2019robustness}). 
    } 
    \label{tab:RQ2}
    
    \vspace{-0.2in}
\end{figure}

\begin{figure*}
    \centering
    \scalebox{0.95}{
        \begin{tabular}{m{0.24\textwidth} m{0.24\textwidth}m{0.24\textwidth}m{0.24\textwidth}}
        
         \begin{subfigure}{.23\textwidth}\captionsetup{justification=centering} \includegraphics[width=\linewidth]{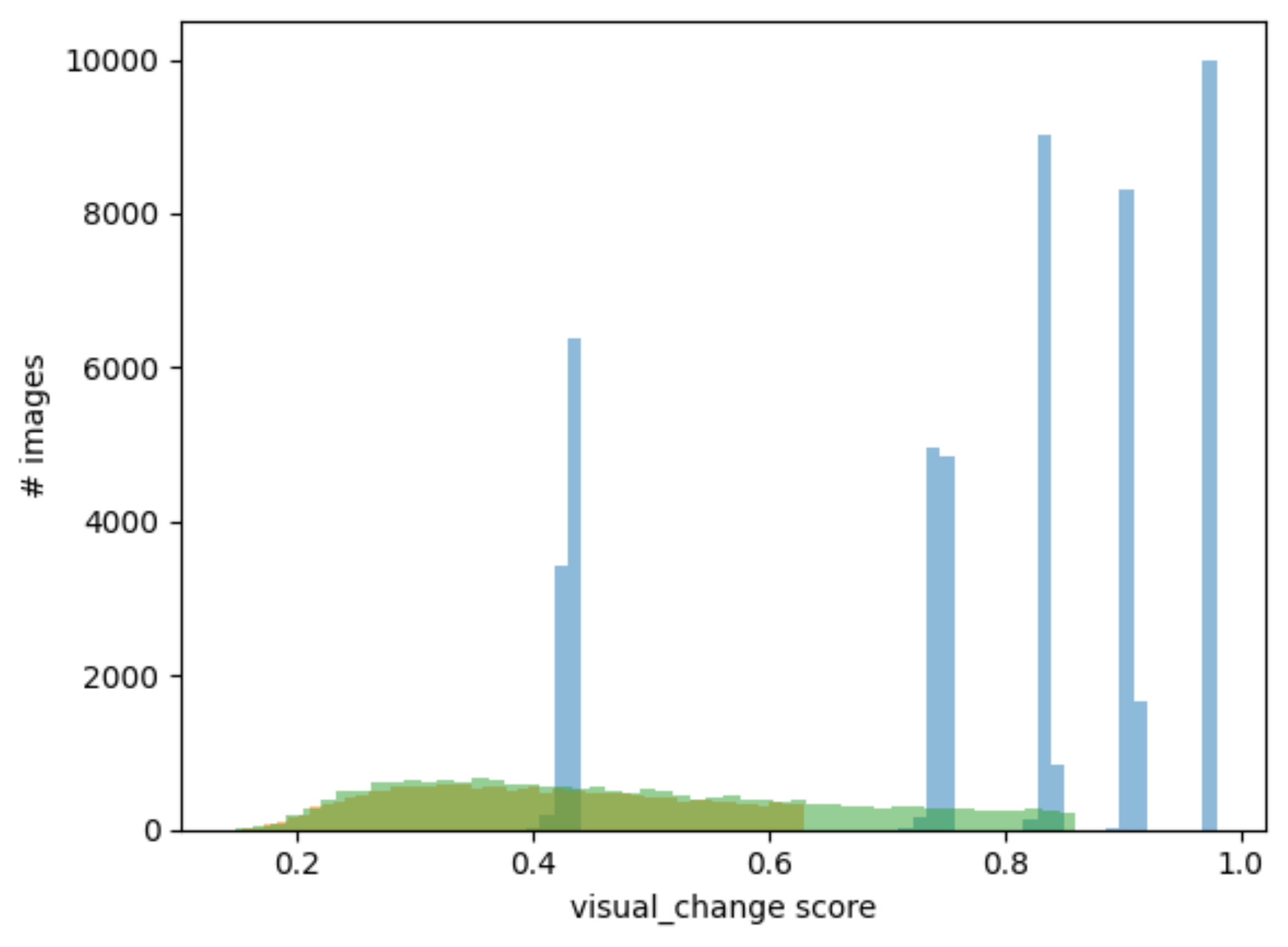}\caption{ Contrast with \\CIFAR-10 images}\label{RQ2_contrast} \end{subfigure} 
            & \begin{subfigure}{.23\textwidth}\captionsetup{justification=centering}\includegraphics[width=\linewidth]{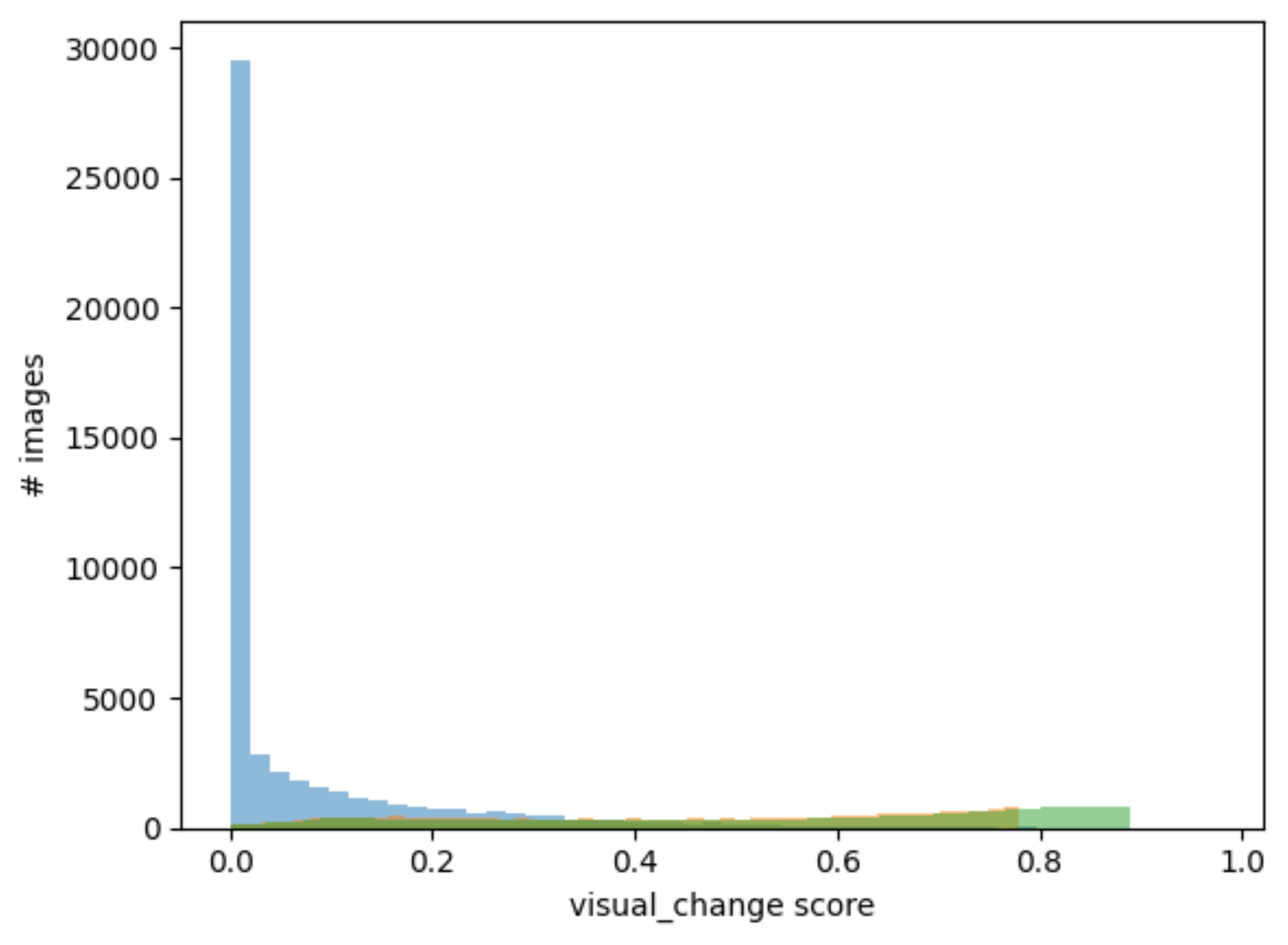}\caption{Brightness with\\CIFAR-10 images}\label{RQ2_brightness} \end{subfigure}  
            & \begin{subfigure}{.23\textwidth}\captionsetup{justification=centering}\includegraphics[width=\linewidth]{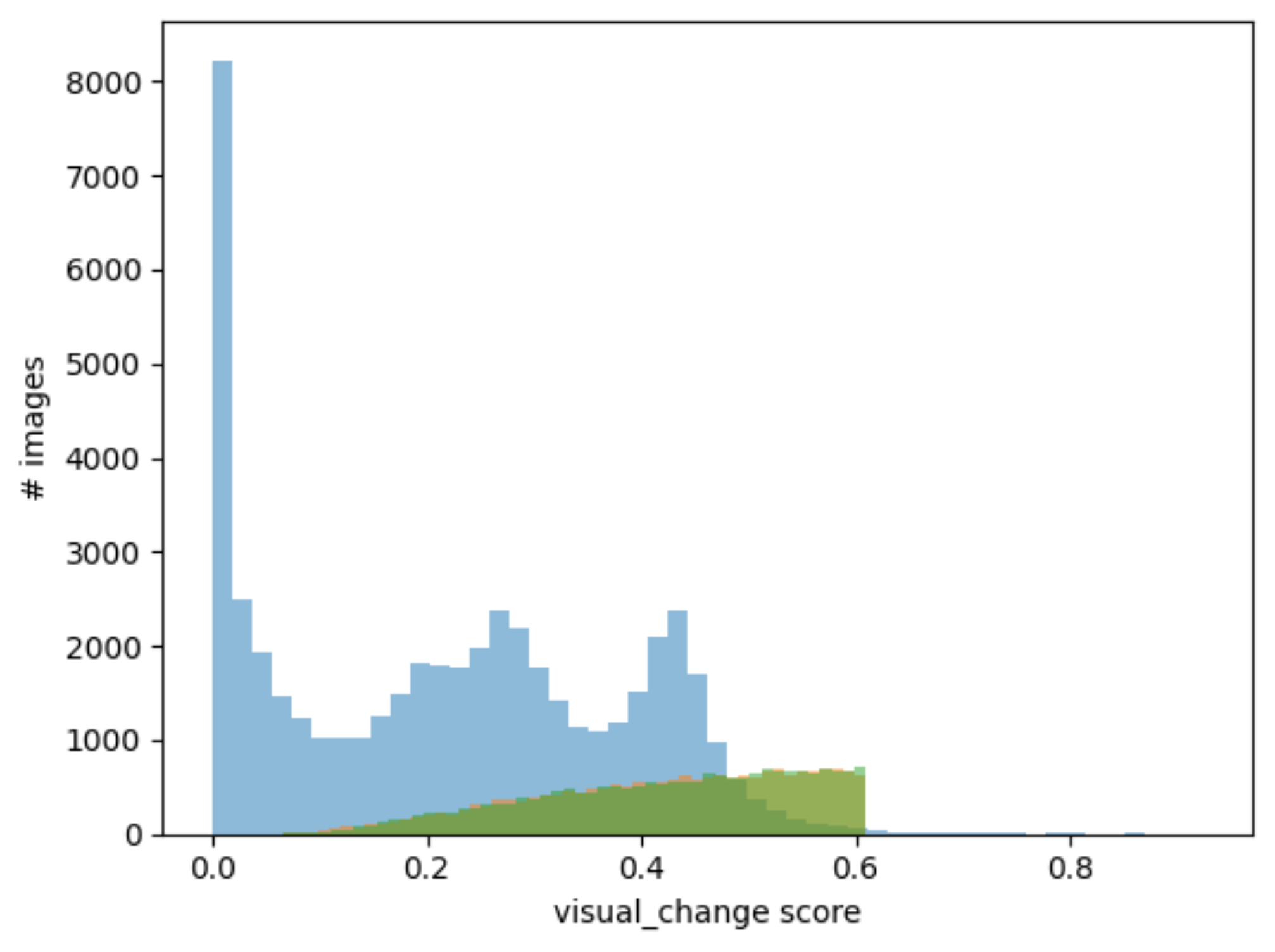}\caption{Frost with \\CIFAR-10 images}\label{RQ2_frost} \end{subfigure} 
            &
            \begin{subfigure}{.23\textwidth}\captionsetup{justification=centering}\includegraphics[width=\linewidth]{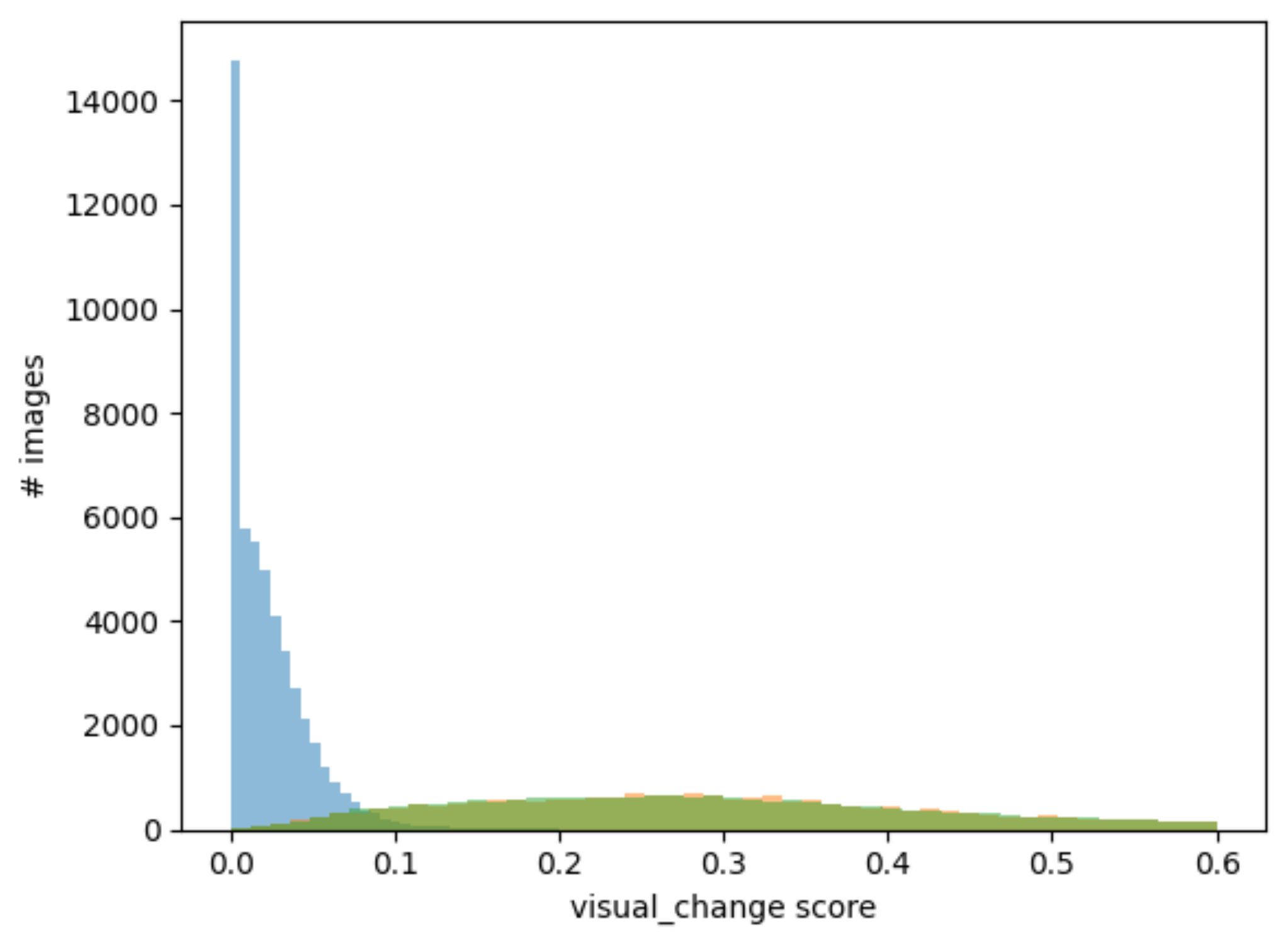}\caption{
                JPEG Compression with \\CIFAR images
            }
            \label{RQ2_compression} 
            \end{subfigure} 
            \\
            \begin{subfigure}{.23\textwidth}\captionsetup{justification=centering}\includegraphics[width=\linewidth]{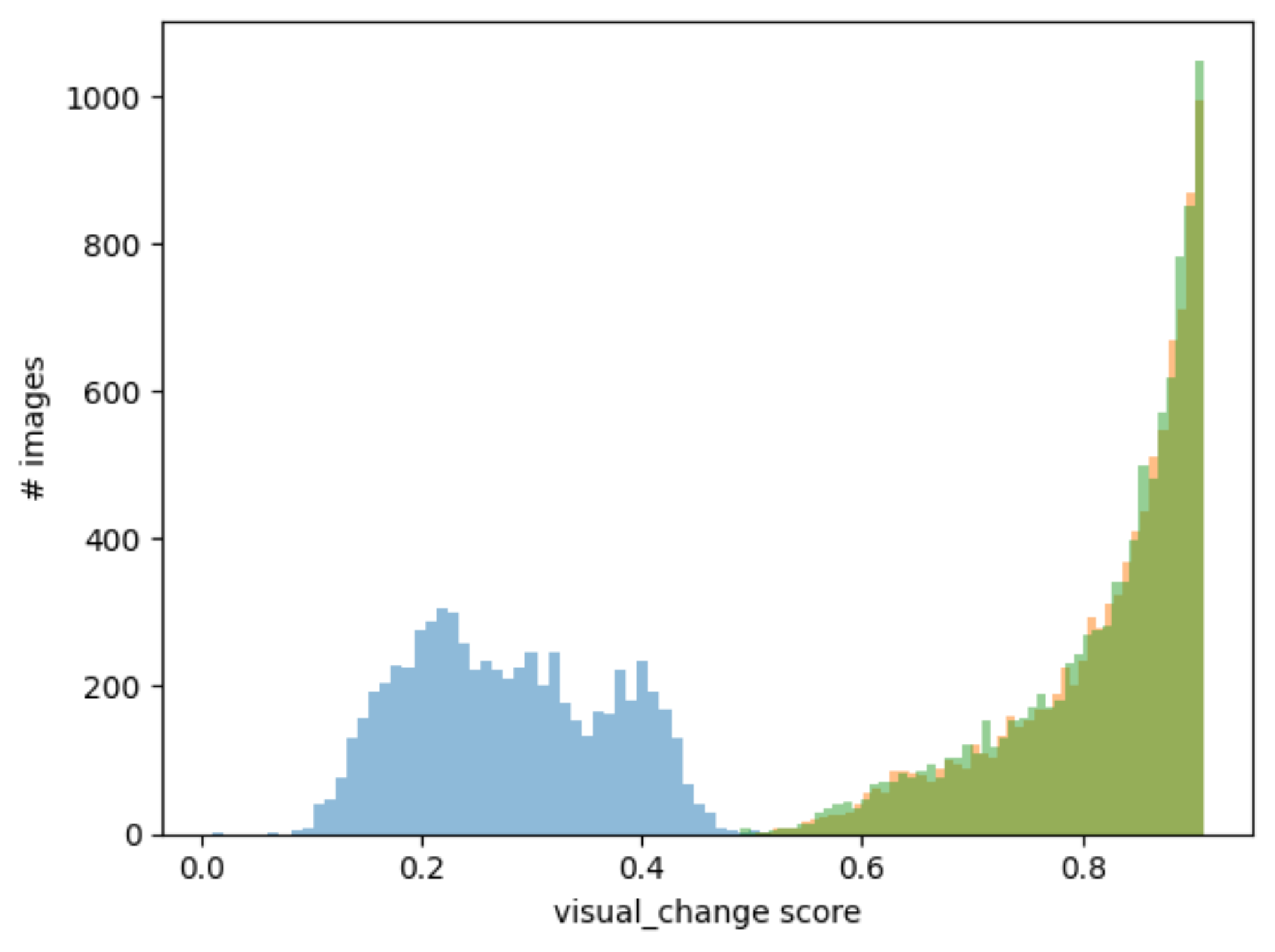}\caption{ Gaussian noise with \\ Imagenet (ILSVRC'12)\\ images}\label{RQ2_contrast_imagenet} \end{subfigure} 
            & \begin{subfigure}{.23\textwidth}\captionsetup{justification=centering}\includegraphics[width=\linewidth]{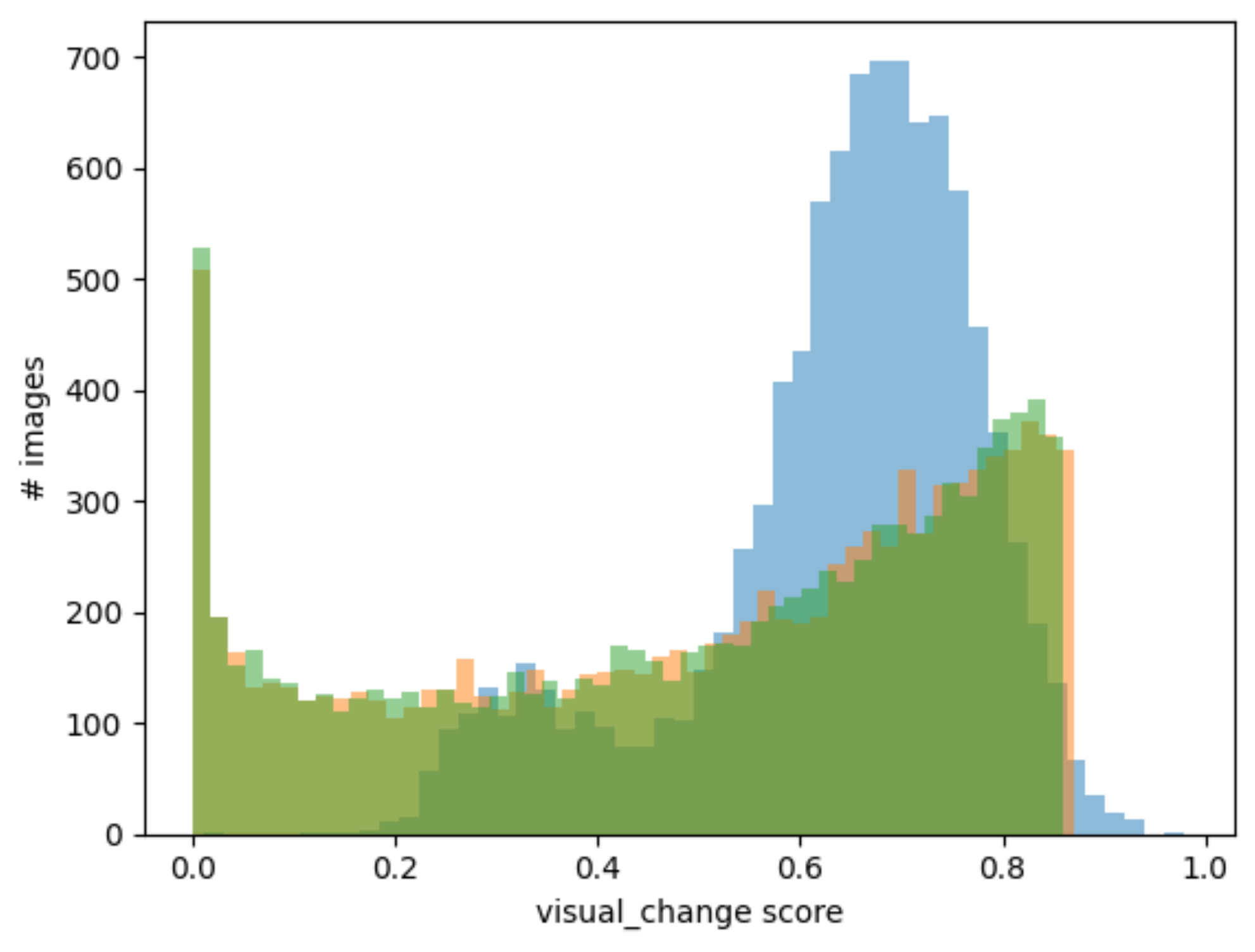}\caption{ Brightness with \\ Imagenet (ILSVRC'12)\\images}\label{RQ2_brightness_imagenet} \end{subfigure} 
            & \begin{subfigure}{.23\textwidth}\captionsetup{justification=centering}\includegraphics[width=\linewidth]{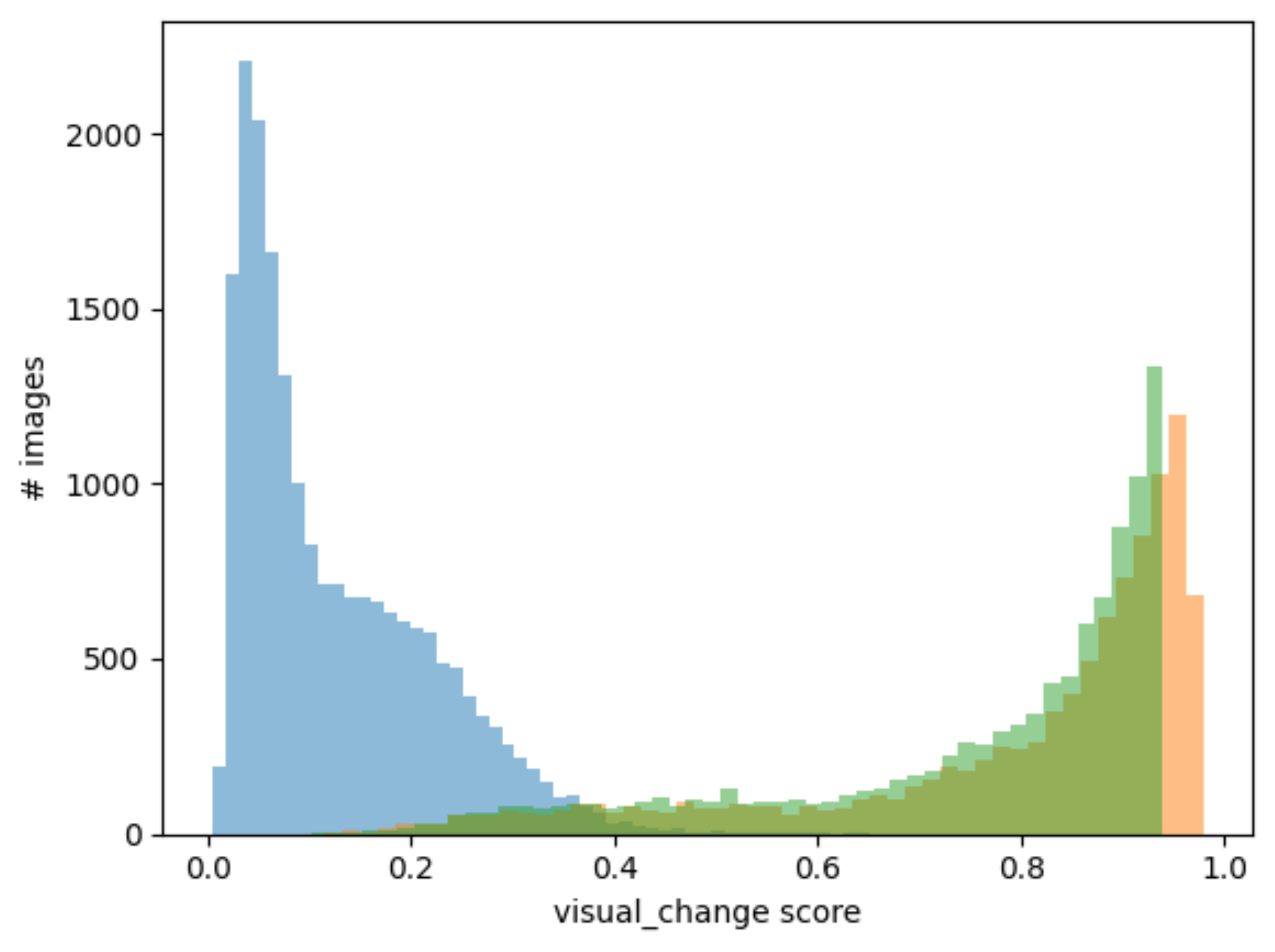}\caption{ Defocus blur with \\ Imagenet (ILSVRC'12)\\images}\label{RQ2_defocus_blur_imagenet} \end{subfigure} 
            &
             \begin{subfigure}{.23\textwidth}\captionsetup{justification=centering}\includegraphics[width=\linewidth]{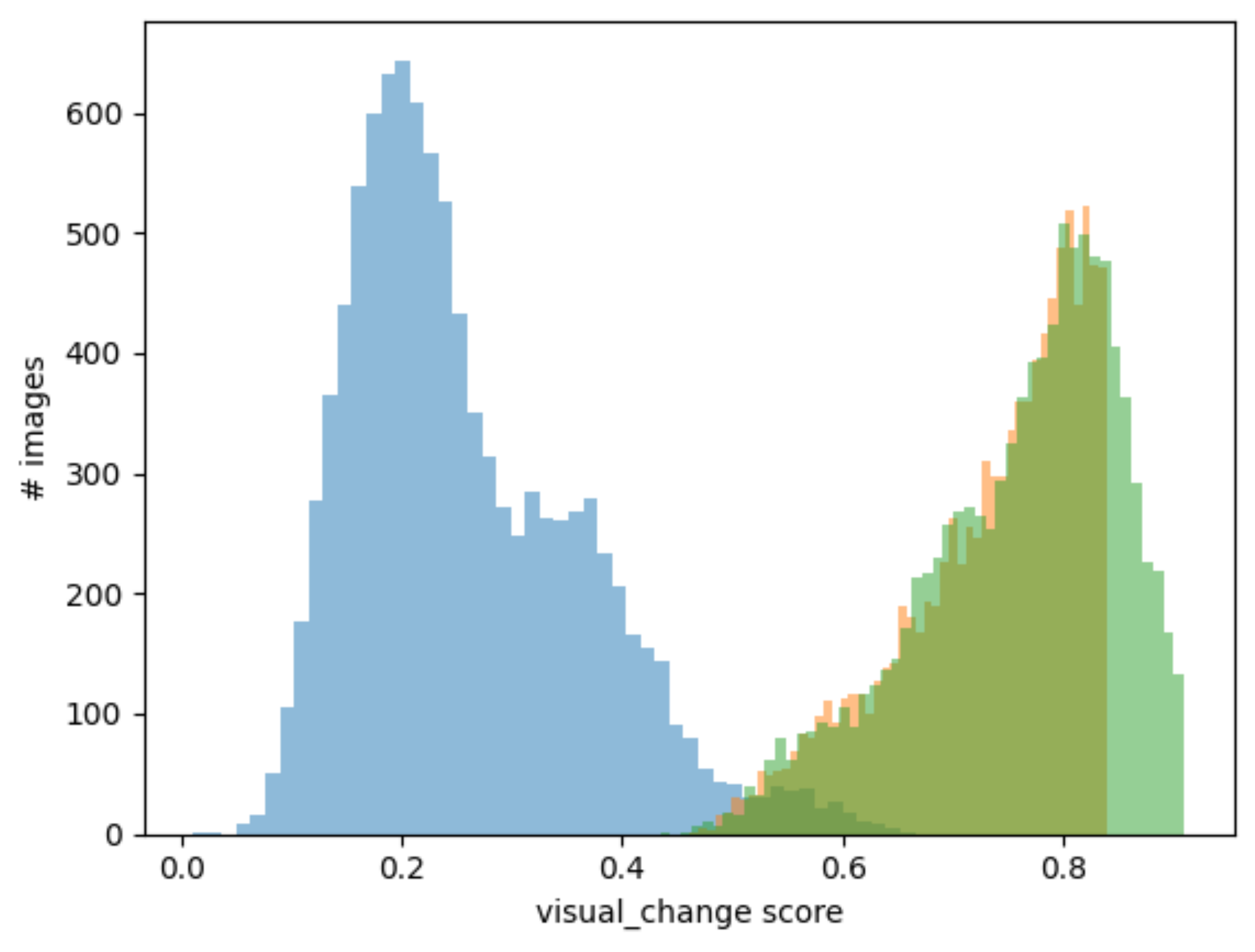}\caption{ Frost with \\Imagenet (ILSVRC'12)\\ images}\label{RQ2_frost_imagenet} \end{subfigure}\\
             \hline
        \multicolumn{4}{|c|}{\textbf{Legend:} \includegraphics[scale=0.5]{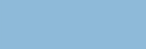} CIFAR-10-c/Imagenet-c images, \includegraphics[scale=0.5]{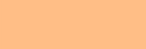} images generated within $[0, t_c]$, \includegraphics[scale=0.5]{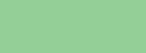} images generated within $[0, t_p]$}\\
        \hline
        \end{tabular}
    }
    \caption{\small A comparison of the range and distribution of $\Delta_v$ scores in test images of robustness benchmark datasets (CIFAR-10-c and Imagenet-c~\cite{hendrycks2019robustness}) and test images generated with our requirement checking method. The x-axis is the $\Delta_v$ score. The y-axis is the number of images. 
    }
    \label{fig:RQ2_plots}
\end{figure*}

To answer RQ2, we first compare our human-tolerated ranges with the ranges of changes included in the robustness benchmark datasets, to see whether existing methods already cover them. In Fig.~\ref{tab:RQ2}, we show, for each transformation, the range of changes in images included in Imagenet-c/ CIFAR-10-c~\cite{hendrycks2019robustness}\footnote{Note that due to the large size of Imagenet-c, the distribution is obtained by uniformly sampling the entire benchmarking dataset.} in blue, and our human-tolerated ranges for both requirements in yellow and green.
The overlapping of ranges indicates the degree to which our ranges are covered by Imagenet-c/ CIFAR-10-c.  The ranges in Imagenet-c and CIFAR-10-c~\cite{hendrycks2019robustness} are either larger (e.g., brightness and frost for CIFAR-10-c; brightness for Imagenet-c) or smaller (e.g., contrast and jpeg compression of CIFAR-10-c; Gaussian noise, defocus blur and frost for Imagenet-c) than the human-tolerated range.
The images included in Imagenet-c/CIFAR-10-c are transformed by using a pre-selected list of five parameter values per transformation. This result shows that simply generating images this way does not address the full range of realistic changes that do not affect human performance. 
Secondly, we compare the distribution of the test cases (transformed images) within the human-tolerated range generated from our requirement checking method and from CIFAR-10-c and Imagenet-c.
Our requirement checking method for generating test cases samples the parameter space uniformly and then transforms the images. As the number of parameters for a transformation increases, so does the possible number of combinations of parameter values that can lead to the same degree of visual change in the images. Therefore, sampling the parameter space uniformly allows us a better coverage of possible transformed images resulting in a fairer reliability evaluation compared with  transformations with pre-selected parameter values, as done in CIFAR-10-c and Imagenet-c~\cite{hendrycks2019robustness}. In Fig.~\ref{fig:RQ2_plots}, we show the distributions of transformed images generated with our requirement checking method and images in CIFAR-10-c and Imagenet-c.
As we can observe from the plots, the transformed images from CIFAR-10-c and Imagenet-c either favor certain ranges of $\Delta_v$ score (Fig.~\ref{RQ2_brightness}, \ref{RQ2_frost}, \ref{RQ2_compression}) or are discontinuous (Fig.~\ref{RQ2_contrast}) and therefore biased. 
This also suggests that the approach of generating transformed images in the benchmark datasets does not guarantee a fair evaluation of reliability within the human-tolerated range of changes because of the biased distribution of tests. 
Thus, the  human-tolerated ranges of changes are not addressed or properly tested by existing methods, answering RQ2.

\vskip 0.05in
\noindent
{\bf RQ3.}
To answer RQ3, we aim to determine whether checking our requirements enables us to  discover reliability gaps that were not identified with existing reliability benchmarks.
We evaluated the reliability of 13 state-of-the-art image classification models with the vision task of recognizing cars in images using both our requirement checking method and the existing benchmarks CIFAR-10-c and Imagenet-c.
Results are shown in Tbl.~\ref{tab:RQ3}. Note that no models satisfy our requirements with $95\%$ confidence, which is not surprising since they were not trained with data covering the human-tolerated range. However, several models pre-trained on Imagenet (ILSVRC’12) images have negative reliability distance ($\hat{s_0}-\hat{s_{t_p}}$) for our \emph{prediction-preservation} requirements, which suggests that these models are close to satisfying these requirements.

For each transformation, the models in Tbl.~\ref{tab:RQ3} are ranked based on the evaluation results (accuracy) of benchmark images.   A higher ranking means that the model is more reliable.
We compare the reliability ranking of these models using our reliability distance for both of our requirements (see Sec.~\ref{testing}) with the benchmark ranking, indicating the differences in blue.
The tests included in the CIFAR-10-c and Imagenet-c benchmarks~\cite{hendrycks2019robustness} are biased toward images with small transformation magnitudes, resulting in significant differences with our ranking for brightness (Fig.~\ref{RQ2_brightness}), frost (Fig.~\ref{RQ2_frost} and~\ref{RQ2_frost_imagenet}), jpeg compression (Fig.~\ref{RQ2_compression}) and defocus blur (Fig.~\ref{RQ2_defocus_blur_imagenet}) transformations.
Therefore, if a model is lower on the ranking of our reliability distance than on the benchmark ranking, it is less reliable than predicted by the benchmark, meaning that our method discovered a new reliability gap. 
Below we summarize the main reliability gaps identified by our method.
(i) RLAT is ranked by CIFAR-10-c within the three last models for the transformations contrast, brightness, and frost, but the first for jpeg compression. However, our method ranks RLAT at the bottom for all the transformations including jpeg compression, indicating that the tests generated by our method 
are able to detect the reliability gap missed by the benchmark.
(ii) For jpeg compression, RLATAugMixNoJSD is ranked second  both by the CIFAR-10-c benchmark and by our correctness-preservation reliability distance. However, RLATAugMixNoJSD is ranked last by our prediction-preservation reliability distance. Similarly, for Gaussian noise, resnext101\_a+d is ranked first by both our \emph{correction-preservation} and Imagenet-c benchmark, but it is ranked second last by our \emph{prediction-preservation} reliability distance. This shows that both RLATAugMixNoJSD and resnext101\_a+d have a high accuracy for transformed images but their predictions are not consistent. 
Therefore, checking our prediction-preservation requirement enabled us to identify new reliability gaps that could not be detected by only checking accuracy on transformed images, answering RQ3.

\input{big_table_new}

\vskip 0.05in
\noindent
{\bf Summary.} Through answering RQ1, we show that parameters of the requirements estimated with our requirement instantiation method are \emph{reusable} for different images sharing the same class and images distribution. Through answering RQ2, we show that existing work does not adequately cover the range of changes that do not affect humans. Finally, through answering RQ3, we show that our requirement checking method is useful, since it can discover reliability gaps that are missed by the existing methods. 
Also, notice that our \emph{prediction-preservation} is close to being satisfied by several models pre-trained on Imagenet (ILSVRC’12) images; this indicates that our requirements are satisfiable. 
Thus, our evaluation suggests that the proposed requirements are useful and reusable for checking reliability of MVCs.

\vskip 0.05in
\noindent
{\bf Threats to validity.}
[Construct] For the correctness-preservation requirement, the human performance may seem too hard for MVCs to match. However, following guidelines provided by Firestone~\cite{Firestone26562}, we choose to keep the requirements for a fair comparison between a human and an ML performance. Further, training with data augmentation that covers the range of visual changes for each transformation as per our requirements might enable an MVC to meet them. Checking this hypothesis is future work. [Internal] We assumed that the parameter values for any transformation should be uniformly distributed. This may be different depending on the application of the MVC, e.g., heavy snow may be less relevant for autonomous cars deployed in tropical regions than other regions. 
[External] Due to budget considerations, we included a limited set of transformations and image classes in our experiments with humans. Experiments for other visual tasks are also future work.

%% file: big_table_new.tex
\pdfoutput=1
\begin{table*}
\small
\centering
 \caption{ {\small A comparison of reliability evaluation of MVCs using our method and using state-of-the-art benchmarks (CIFAR-10-c and Imagenet-c).
  For each transformation and the visual task of car recognition, the MVC models are ranked w.r.t. their accuracy on all benchmark images.  
  $\hat{m}_0$ and $\hat{s}_0$ are, resp., the required accuracy and percentage of labels preserved in our requirements. $\hat{m}_{t_c}$ and $\hat{s}_{t_p}$ are, resp., the resulting accuracy and perception preservation percentage through checking the models against our requirements.
  The differences between the benchmark ranking and our ranking using the reliability distance are highlighted in blue.
}}
  
  \scalebox{0.72}{
  \begin{tabular}{|c|c|c|c|c|c|c|c|c|c|c|c|c|c|}
    \hline
     \cellcolor{gray!25} &\cellcolor{gray!25}& \cellcolor{gray!25}& \multicolumn{2}{c}{\cellcolor{gray!25}\shortstack{Checking \textbf{our} \\Correctness-preservation}} & \multicolumn{2}{|c|}{\cellcolor{gray!25}\shortstack{Checking \textbf{our} \\Prediction-preservation }} &\cellcolor{gray!25} &\cellcolor{gray!25} & \multicolumn{2}{c}{\cellcolor{gray!25}\shortstack{Checking \textbf{our} \\Correctness-preservation }} & \multicolumn{2}{|c|}{\cellcolor{gray!25}\shortstack{Checking \textbf{our} \\Prediction-preservation}}\\
      \cline{4-7} \cline{10-13}
       \multirow{-5}{*}{\cellcolor{gray!25}\rotatebox[origin=c]{90}{dataset}}&\multirow{-4}{*}{\cellcolor{gray!25}\shortstack{model name}} & \multirow{-6}{*}{\cellcolor{gray!25}\shortstack{accuracy \\on all \\\textbf{benchmark}\\images}} &\cellcolor{gray!25}  \shortstack{required\\ and \\ estimated \\accuracy\\ $\hat{m}_0|\hat{m}_{t_c}$} & \cellcolor{gray!25}\shortstack{reliability\\distance \\ $\hat{m}_0-\hat{m}_{t_c}$ \\(rank)}  &\cellcolor{gray!25} \shortstack{required\\ and \\ estimated \\percentage\\ $\hat{s}_0| \hat{s}_{t_p}$} & \cellcolor{gray!25}\shortstack{reliability \\distance \\ $\hat{s}_0-\hat{s}_{t_p}$ \\ (rank) } &\multirow{-4}{*}{\cellcolor{gray!25}\shortstack{model name}} & \multirow{-6}{*}{\cellcolor{gray!25}\shortstack{accuracy \\on all \\\textbf{benchmark}\\images}} &\cellcolor{gray!25}\shortstack{required\\ and \\ estimated \\accuracy\\ $\hat{m}_0|\hat{m}_{t_c}$} &\cellcolor{gray!25}\shortstack{reliability \\distance \\ $\hat{m}_0-\hat{m}_{t_c}$ \\ (rank) }   &\cellcolor{gray!25}\shortstack{required\\ and \\ estimated \\percentage\\ $\hat{s}_0|\hat{s}_{t_p}$} &\cellcolor{gray!25} \shortstack{reliability \\distance \\ $\hat{s}_0-\hat{s}_{t_p}$ \\ (rank)} \\
      \hline
      \parbox[t]{2mm}{\multirow{17}{*}{\rotatebox[origin=c]{90}{\textbf{cifar-10}}}} & \multicolumn{6}{|c}{\cellcolor{gray!10}\shortstack{\textbf{contrast}}} & \multicolumn{6}{|c|}{\cellcolor{gray!10}\shortstack{\textbf{brightness}}} \\
    \cline{2-13}
      & Augmix\_ResNeXt~\cite{hendrycks2020augmix} &0.9920 & \cellcolor{blue!20}0.9961 | 0.9871& \cellcolor{blue!20}(2) 0.009  & 0.996 | \cellcolor{blue!20}0.9761& \cellcolor{blue!20}(2) 0.0199 & Augmix\_ResNeXt~\cite{hendrycks2020augmix} & 0.9952& \cellcolor{blue!20}0.9963 | 0.9776 & \cellcolor{blue!20}(3) 0.0187  &\cellcolor{blue!20}0.999 | 0.9576 & \cellcolor{blue!20}(3) 0.0414 \\
      \cline{2-13}
     & Augmix\_WRN~\cite{hendrycks2020augmix} & 0.9909& \cellcolor{blue!20}0.9952 | 0.9868& \cellcolor{blue!20}(1) 0.0084  &  \cellcolor{blue!20}0.995 | 0.9756 & \cellcolor{blue!20}(1) 0.0194  &  AugMixNoJSD~\cite{kireev2021effectiveness} & 0.9945 & \cellcolor{blue!20}0.9961 | 0.9782 & \cellcolor{blue!20}(1) 0.0179 & \cellcolor{blue!20}0.996 | 0.9573& \cellcolor{blue!20}(1) 0.0387\\
      \cline{2-13}
     & AugMixNoJSD~\cite{kireev2021effectiveness} & 0.9901& 0.9952 | 0.9851 & (3) 0.0101 & 0.997 | 0.9674 & (3) 0.0296 & \shortstack{Augmix\_WRN~\cite{hendrycks2020augmix}} & 0.9943& \cellcolor{blue!20}0.9953 | 0.9768  & \cellcolor{blue!20}(2) 0.0185  & \cellcolor{blue!20}0.994 | 0.9540 &  \cellcolor{blue!20}(2) 0.04\\
      \cline{2-13}
      & Standard~\cite{Zagoruyko2016WideRN} & 0.9862 & 0.9952 | 0.9809 & (4) 0.0143  & 0.994 | 0.9570 & (4) 0.037 & \shortstack{RLATAugMixNoJSD~\cite{kireev2021effectiveness}} & 0.9942& 0.9953 | 0.9730 & (4) 0.0223  & 0.998 | 0.9488 & (4) 0.0492 \\
      \cline{2-13}
     & \shortstack{RLATAugMixNoJSD~\cite{kireev2021effectiveness}} & 0.9788 &  0.9936 | 0.9710  & (5) 0.0226 & 0.993 | 0.9416 & (5) 0.0514  & Standard~\cite{Zagoruyko2016WideRN} &0.9933& 0.9947 | 0.9690& (5) 0.0257 & 0.998 | 0.9451& (5) 0.0529\\
      \cline{2-13}
      & Gauss50percent~\cite{kireev2021effectiveness} & 0.9577 & 0.9925 | 0.9261 & (6) 0.0664  &0.987 | 0.8994& (6) 0.0876 &  RLAT~\cite{kireev2021effectiveness} & 0.9928& \cellcolor{blue!20}0.9930 | 0.9557& \cellcolor{blue!20}(7) 0.0373 & 0.993 | 0.9307 & (6) 0.0623 \\
      \cline{2-13}
      &RLAT~\cite{kireev2021effectiveness} & 0.9550 & 0.9936 | 0.9133 &  (7) 0.0803  & 0.991 | 0.8880 & (7) 0.103  & Gauss50percent~\cite{kireev2021effectiveness} & 0.9904& \cellcolor{blue!20}0.9925 | 0.9555& \cellcolor{blue!20}(6) 0.037 & 0.995 | 0.9305 & (7) 0.0645\\
      \cline{2-13}

     \cline{2-13}
     & \multicolumn{6}{|c}{\cellcolor{gray!10}\shortstack{\textbf{frost}}} & \multicolumn{6}{|c|}{\cellcolor{gray!10}\shortstack{\textbf{JPEG compression}}} \\
     
     \cline{2-13}
    & Augmix\_ResNeXt~\cite{hendrycks2020augmix} & 0.9912 & \cellcolor{blue!20}0.9969 | 0.9771 &  \cellcolor{blue!20}(2) 0.0198  & 0.995 | 0.9776 & (1) 0.0174  & RLAT~\cite{kireev2021effectiveness} & 0.9910 & \cellcolor{blue!20}0.9927 | 0.9443 & \cellcolor{blue!20}(5) 0.0484  & 0.999 | 0.9773 & (1) 0.0217 \\
     \cline{2-13}
     & \shortstack{RLATAugMixNoJSD~\cite{kireev2021effectiveness}} & 0.9910 & \cellcolor{blue!20}0.9958 | 0.9737 & \cellcolor{blue!20}(4) 0.0221  & 0.994 | 0.9738 &(2) 0.0202  & \shortstack{RLATAugMixNoJSD~\cite{kireev2021effectiveness}}  & 0.9899 & \cellcolor{blue!20}0.9942 | 0.9659 & \cellcolor{blue!20}(2) 0.0283 & \cellcolor{blue!20}0.999 | 0.9365 & \cellcolor{blue!20}(7) 0.0625 \\
     \cline{2-13}
     & Augmix\_WRN~\cite{hendrycks2020augmix} &0.9899 & \cellcolor{blue!20}0.9955 | 0.9765 &  \cellcolor{blue!20}(1) 0.019& \cellcolor{blue!20}0.998 | 0.9758 & \cellcolor{blue!20}(4) 0.0222  &  Gauss50percent~\cite{kireev2021effectiveness} & 0.9897 & \cellcolor{blue!20}0.9915 | 0.9701 & \cellcolor{blue!20}(1) 0.0214  & \cellcolor{blue!20}0.999 | 0.9735 &  \cellcolor{blue!20}(2) 0.0255 \\
     \cline{2-13}
   &  AugMixNoJSD~\cite{kireev2021effectiveness} & 0.9890 & \cellcolor{blue!20}0.9965 | 0.9754 &  \cellcolor{blue!20}(3) 0.0211 &\cellcolor{blue!20}0.997 | 0.9754 & \cellcolor{blue!20}(3) 0.0216  & Augmix\_ResNeXt~\cite{hendrycks2020augmix}  & 0.9894 & \cellcolor{blue!20}0.9949 | 0.9516 & \cellcolor{blue!20}(4) 0.0433 & \cellcolor{blue!20}0.999 | 0.9529& \cellcolor{blue!20}(4) 0.0461 \\
     \cline{2-13}
      & RLAT~\cite{kireev2021effectiveness} & 0.9875 & \cellcolor{blue!20}0.9948 | 0.9414 &  \cellcolor{blue!20}(7) 0.0534  &\cellcolor{blue!20}0.986 | 0.9430 & \cellcolor{blue!20}(7) 0.043  & Augmix\_WRN~\cite{hendrycks2020augmix} & 0.9886 & \cellcolor{blue!20}0.9942 | 0.9511 & \cellcolor{blue!20}(3) 0.0431 & \cellcolor{blue!20}0.999 | 0.9547 & \cellcolor{blue!20} (3) 0.0443 \\
      \cline{2-13}
     & Gauss50percent~\cite{kireev2021effectiveness} & 0.9867 & \cellcolor{blue!20}0.9933 | 0.9506 &  \cellcolor{blue!20}(6) 0.0427  & \cellcolor{blue!20}0.99 | 0.9524 & \cellcolor{blue!20}(5) 0.0376  & AugMixNoJSD~\cite{kireev2021effectiveness} & 0.9868 & 0.9953 | 0.9443 & (6) 0.051  & \cellcolor{blue!20}0.998 | 0.9471 &  \cellcolor{blue!20} (5) 0.0509 \\
      \cline{2-13}
      &Standard~\cite{Zagoruyko2016WideRN} & 0.9752 & \cellcolor{blue!20}0.9956 | 0.9567 &  \cellcolor{blue!20}(5) 0.0389 & \cellcolor{blue!20}0.997 | 0.9564 & \cellcolor{blue!20}(6) 0.0406 & Standard~\cite{Zagoruyko2016WideRN} & 0.9734& 0.9952 | 0.9359 & (7) 0.0593  & \cellcolor{blue!20}0.996 | 0.9365 & \cellcolor{blue!20}(6) 0.0595 \\
      \hline
    \hline
    
     \parbox[t]{2mm}{\multirow{12}{*}{\rotatebox[origin=c]{90}{\textbf{imagenet}}}} &
      \multicolumn{6}{|c}{\cellcolor{gray!10}\textbf{Gaussian noise}} & \multicolumn{6}{|c|}{\cellcolor{gray!10}\textbf{frost}} \\
     
     \cline{2-13}
    &  resnext101\_a+d~\cite{Hendrycks2020TheMF} &0.9962 & 0.997 | 0.9959 & (1) 0.0011 & \cellcolor{blue!20}0.998 | 0.997& \cellcolor{blue!20}(5) 0.001  & resnext101\_a+d~\cite{Hendrycks2020TheMF} & 0.9958& 0.9974 | 0.9954 & (1) 0.002 &\cellcolor{blue!20}0.996 | 0.9962 & \cellcolor{blue!20}(1) \textbf{-0.0002} \\
     \cline{2-13}
    & aug+deep~\cite{Hendrycks2020TheMF}  & 0.9956 & 0.9958 | 0.9942 & (2) 0.0016  & \cellcolor{blue!20}0.996 | 0.9961 & \cellcolor{blue!20}(1) \textbf{-0.0001}  &  aug+deep~\cite{Hendrycks2020TheMF} & 0.9952 & 0.9967 | 0.9943 & (2) 0.0024 &\cellcolor{blue!20} 0.996 | 0.9942 & \cellcolor{blue!20}(5) 0.0018\\
      \cline{2-13}
    &  deepaugment~\cite{Hendrycks2020TheMF} & 0.9955 & \cellcolor{blue!20}0.9963 | 0.9937 & \cellcolor{blue!20}(4) 0.0026 & \cellcolor{blue!20}0.996 | 0.9959 & \cellcolor{blue!20}(3) 0.0001  & ANT3x3\_SIN~\cite{Rusak2020IncreasingTR} & 0.9944& 0.996 | 0.9935  & (3) 0.0025  &\cellcolor{blue!20} 0.992 | 0.9924 &  \cellcolor{blue!20}(1) \textbf{-0.0004} \\
      \cline{2-13}
    &  ANT\_SIN~\cite{Rusak2020IncreasingTR} & 0.9946 & \cellcolor{blue!20}0.9953 | 0.9935 & \cellcolor{blue!20}(3) 0.0018 & \cellcolor{blue!20}0.996 | 0.9962 & \cellcolor{blue!20}(1) \textbf{-0.0002} & ANT\_SIN~\cite{Rusak2020IncreasingTR} & 0.9941& 0.9962 | 0.9927 & (4) 0.0035  & \cellcolor{blue!20}0.99 | 0.9927 &\cellcolor{blue!20} (1) \textbf{-0.0027} \\
     \cline{2-13}
    & Speckle\_Model~\cite{Rusak2020IncreasingTR} & 0.9934 &  0.9958 | 0.9916  & (5) 0.0042 & \cellcolor{blue!20}0.996 | 0.9952 & \cellcolor{blue!20}(4) 0.0008  & deepaugment~\cite{Hendrycks2020TheMF} & 0.9936 & 0.9966 | 0.993& (5) 0.0036  & \cellcolor{blue!20}0.994 | 0.992 & \cellcolor{blue!20}(6) 0.002 \\
      \cline{2-13}
   & resnet50~\cite{Resnet50} & 0.9924 &  0.9953 | 0.9908  & (6) 0.0045  & 0.996 | 0.9949 & (6) 0.0011  & resnet50~\cite{Resnet50} & 0.9921 & 0.9957 | 0.9907& (6) 0.005  & \cellcolor{blue!20}0.992 | 0.9911 & \cellcolor{blue!20}(4) 0.0009 \\
     \cline{2-13}
    & \multicolumn{6}{|c}{\cellcolor{gray!10}\shortstack{\textbf{brightness}}} & \multicolumn{6}{|c|}{\cellcolor{gray!10}\shortstack{\textbf{defocus blur}}} \\
     
     \cline{2-13}
    &  resnext101\_a+d~\cite{Hendrycks2020TheMF} & 0.9972 & \cellcolor{blue!20}0.9967 | 0.9953 & \cellcolor{blue!20} (2) 0.0014  & \cellcolor{blue!20}1 | 0.9972 & \cellcolor{blue!20}(4) 0.0028  & resnext101\_a+d~\cite{Hendrycks2020TheMF} & 0.9949 & 0.9977 | 0.995 & (1) 0.0027 & 0.994 | 0.9957 &  (1) \textbf{-0.0017} \\
      \cline{2-13} 
   &   aug+deep~\cite{Hendrycks2020TheMF} & 0.9966 & \cellcolor{blue!20}0.9959 | 0.9947 &  \cellcolor{blue!20}(1) 0.0012& \cellcolor{blue!20}1 | 0.9964 & \cellcolor{blue!20}(5) 0.0036  & aug+deep~\cite{Hendrycks2020TheMF}  & 0.9946 & 0.9972 | 0.9937 & (2) 0.0035  & 0.996 | 0.9943 & (2) 0.0017 \\
     \cline{2-13}
    &  deepaugment~\cite{Hendrycks2020TheMF} & 0.9959 &\cellcolor{blue!20}0.9956 | 0.9937 &\cellcolor{blue!20} (3) 0.0019 & \cellcolor{blue!20}0.996 | 0.9949 & \cellcolor{blue!20}(2) 0.0011& deepaugment~\cite{Hendrycks2020TheMF} & 0.9924 & \cellcolor{blue!20}0.9965 | 0.9914 & \cellcolor{blue!20}(5) 0.005 & \cellcolor{blue!20}0.998 | 0.9929 &  \cellcolor{blue!20}(5) 0.0051 \\
    \cline{2-13}
     & ANT3x3\_SIN~\cite{Rusak2020IncreasingTR} & 0.9957 & \cellcolor{blue!20}0.9954 | 0.9926 &  \cellcolor{blue!20}(5) 0.0028 &\cellcolor{blue!20}0.996 | 0.994 &\cellcolor{blue!20} (3) 0.002  & ANT\_SIN~\cite{Rusak2020IncreasingTR} & 0.9920 & \cellcolor{blue!20}0.997 | 0.9917 & \cellcolor{blue!20}(6) 0.053  & \cellcolor{blue!20}0.998 | 0.9929 &  \cellcolor{blue!20}(5) 0.0051\\
      \cline{2-13}
    &  ANT\_SIN~\cite{Rusak2020IncreasingTR} & 0.9956 &\cellcolor{blue!20}0.9954 | 0.993 & \cellcolor{blue!20} (4) 0.0024  &\cellcolor{blue!20}0.993 | 0.998 & \cellcolor{blue!20}(1) \textbf{-0.005} & ANT3x3\_SIN~\cite{Rusak2020IncreasingTR}  & 0.9919 & \cellcolor{blue!20}0.9963 | 0.9924 & \cellcolor{blue!20}(3) 0.0036  &\cellcolor{blue!20} 0.998 | 0.9931 &  \cellcolor{blue!20}(4) 0.0049 \\
    \cline{2-13}
    & resnet50~\cite{Resnet50} & 0.995 &  0.9956 | 0.9917  & (6) 0.0039  & 1 | 0.9937 & (6) 0.0063  & resnet50~\cite{Resnet50} & 0.9909 & \cellcolor{blue!20}0.9961 | 0.9921& \cellcolor{blue!20}(4) 0.0040 & \cellcolor{blue!20}0.996 | 0.9922 & \cellcolor{blue!20}(3) 0.0038 \\
     \hline
      
      \multicolumn{13}{|l|}{\textbf{Note}: Accuracy is calculated with (true positive + true negative) / all images; all the accuracy values are closed to 1 because of the binary classification task. All numbers are rounded.}\\
      \hline
\end{tabular}}
  \label{tab:RQ3}
\end{table*}

%% file: relatedwork.tex
\pdfoutput=1
\vspace{-0.1in}
\section{Related work}
\label{sec:rw}
In this section, we first review the software engineering (SE) approaches defining reliability of MVCs, then the SE and the computer vision (CV) approaches for evaluating reliability of MVCs and, finally, those comparing human performance against MVCs.

\vskip 0.05in
\noindent
{\bf Specifying reliability of MVCs.}
The inductive data-driven nature of machine-learning creates several challenges for requirements specification and verification in MVCs. Yet, multiple recent studies explored this area~\cite{Vogelsang-19,RE2019MLC,superintelligence}.
While they agree on the necessity of requirements elicitation in MVCs, they fail to provide a systematic approach for inferring the requirements.
Several authors attempted to specify the expected behaviour of MVCs indirectly through specifying a set of quality characteristics for training datasets~\cite{medicaldataset2017}, specifying additional ML-related requirements for each phase of software development processes~\cite{salay2018using}, specifying higher-level requirements~\cite{Gauerhof2020}, or specifying how MVCs address the target applications~\cite{seshia2016verified}. 
Yet these approaches cannot be used to check  reliability of MVCs automatically, whereas our reliability requirements are machine-verifiable.

\vskip 0.05in
\noindent
{\bf Checking reliability.}
Metrics for testing MVCs robustness against image transformations  have been defined using metamorphic testing~\cite{metamorphic_bugs, xie2011metatesting, Mekala_2019, zhang2018deeproad}. In contrast, we focus on a different set of transformations, the ones that do degrade the image quality while preserving the human opinion in the image rather than transformations that can be covered with equivariant relations. 
Existing works also use testing to generate corner cases~\cite{deep_validation2019}, or corner case tests to increase MVCs robustness~\cite{Wang2021RobOTRT,ADAPT}.
In contrast, our approach does not focus on corner-cases, but rather on typical cases that can be found in real-world deployments while preserving the human opinion about the content.

Several works evaluated safety of MVCs through assessing their robustness against adversarial examples, either by providing a testing approach to generate adversarial examples~\cite{wicker-18,melis-17,serban-20} in the SE area, providing robustness benchmarks~\cite{dong2020benchmarking,michaelis2019dragon,croce2020robustbench} or verifying the presence of adversarial examples in a given range of image modifications~\cite{huang2017verif, shekar2021label} in the CV area.
In contrast, our focus is on defining boundaries of image modifications using human performance within which the MVCs are expected to maintain their robustness. Also, we do not consider an arbitrary range of image modifications; our approach estimates the range of transformation levels that does not affect human performance.
Previously, we presented the idea of defining adversarial examples using IQA models~\cite{AIRE2020}, focusing only on non-visible changes. In contrast, our current approach considers both visible and non-visible changes in a broader range of real world scenarios.

\vskip 0.05in
\noindent
{\bf Comparing human against machines.}
Prior studies also referred to human performance as the benchmark for the evaluation of their proposed methods~\cite{STALLKAMP2012323}, to better study the existing differences between human and neural networks~\cite{Firestone26562}, to study invariant transformations~\cite{kheradpisheh2016deep}, to compare recognition accuracy~\cite{HoPhuoc2018CIFAR10TC}, or to compare robustness~\cite{Geirhos2018GeneralisationIH}.
In contrast, our focus is not on comparing humans performance with MVCs, but rather on the ranges of transformation magnitudes that do not affect human performance.

%% file: conclusion.tex
\pdfoutput=1
\section{conclusion}
\label{conclusion}
In this paper, we defined reliability of machine vision components (MVC) as `if a human can see it, so should the MVC'.
More precisely, we specified two classes of reliability requirements: correctness-preservation and prediction-preservation. 
Our requirements specify that an MVC should be reliably unaffected by safety-related image transformations, at least within the range of changes that does not affect humans.
We showed, through an evaluation with 13 state-of-the-art pre-trained image classification models, that our approach captures reliability gaps that state-of-the-art reliability methods are unable to detect.
Therefore, we conclude that checking this human tolerated range is important to help software engineers ensure quality and reliability of MVCs. While not discussed in the paper, our requirements can be used for other SE tasks such as 
checking refinement from higher-level system requirements, and checking consistency and compatibility with requirements of other connected components.

In the future, we aim to improve of our  requirement-checking process by providing reliability diagnosis that would help software engineers understand the reliability gaps in their MVCs.
We also aim to validate, through additional experiments, our assumption that our approach can be applicable beyond image classification models, e.g., to handle object detection.  
Finally, we aim to use our reliability requirements for MVCs to provide evidence for building safety assurance cases for the overall system.